# On the three-dimensional spatial correlations of curved dislocation systems

Joseph Pierre Anderson[1]*, Anter El-Azab[1]

[1]School of Materials Engineering

Neil Armstrong Hall of Engineering

701 West Stadium Avenue

West Lafayette, IN 47907-2045

*indicates corresponding author

**Abstract:** Coarse-grained descriptions of dislocation motion in crystalline metals inherently represent a loss of information regarding dislocation-dislocation interactions. In the present work, we consider a coarse-graining framework capable of re-capturing these interactions by means of the dislocation-dislocation correlation functions. The framework depends on a convolution length to define slip-system-specific dislocation densities. Following a statistical definition of this coarse-graining process, we define a spatial correlation function which will allow the arrangement of the discrete line system at two points—and thus the strength of their interactions at short range—to be recaptured into a mean field description of dislocation dynamics. Through a statistical homogeneity argument, we present a method of evaluating this correlation function from discrete dislocation dynamics simulations. Finally, results of this evaluation are shown in the form of the correlation of dislocation densities on the same slip-system. These correlation functions are seen to depend weakly on plastic strain, and in turn, the dislocation density, but are seen to depend strongly on the convolution length. Implications of these correlation functions in regard to continuum dislocation dynamics as well as future directions of investigation are also discussed.

## Introduction

The dislocation-dislocation correlations represent an important link between the continuum and discrete descriptions of the dislocation dynamics. Many views on what this correlation represents, how to evaluate it, and what kinetically-relevant information it contains have been presented in recent years. The present work puts forward a clear and robust definition of the dislocation-dislocation correlation functions and presents a methodology for their computation using simulations of discrete dislocation systems.

One may think of correlation functions as a certain error estimate on mean field representations of discrete systems (cf. self-consistent field theories, Hartree-type theories of electronic systems (Hartree, 1928)). Specifically in our case, the dislocation-dislocation correlation functions represent an error estimate on mean dislocation density field theories (El-Azab et al., 2018). Therefore, to even define a correlation, we must first have some idea of what we are referring to as our mean dislocation density field. Several descriptions have been proposed in recent years, both for the two-dimensional (2D) case of perfectly parallel edge dislocations (Groma, 1997, Groma et al., 1999, Valdenaire et al., 2016), and the three-dimensional (3D) case. In three dimensions, the problem of densities of curved dislocations has been treated by two distinct theories. One which considers a single-valued vector density of dislocations at every point in space (Xia, 2016) and another higher-order theory of curved dislocations which considers many orientations of dislocations at a single point (Hochrainer, 2007, Hochrainer et al., 2014, Sandfeld, 2010). For the purpose of this work, we will consider the former vector density theory of 3D, curved dislocations in face-centered cubic (FCC) crystals by distinguishing each of the 12 slip systems $[\beta]$ as a separate (vector) density field $\boldsymbol{\rho}^{[\beta]}(\boldsymbol{r})$. The first instance of this construction in the literature was due to Anthony (Anthony et al., 1998), and upheld by Kröner (Kröner, 2001) in his final survey on continuum dislocation dynamics. The distinguishing between slip systems , coupled with the high resolution of vector density theories allows one to address a certain insufficiency of the Kröner-Nye tensor, $\boldsymbol{\alpha}(\boldsymbol{r}) \coloneqq \sum_{\beta=1}^{12} \boldsymbol{\rho}^{[\beta]}(\boldsymbol{r}) \otimes \boldsymbol{b}^{[\beta]}$, to predict its own evolution (El-Azab et al., 2018, Hochrainer, 2007, Kröner, 2001). Beyond the separate treatment of slip system densities and the expression of the line direction as a vector valued density, no other quantities are necessary to define the correlations. However, we would like to examine the impact of the spatial resolution on the current corpus of literature on dislocation correlations.

Several researchers have considered the dislocation density as the spatial convolution of the discrete dislocation lines with some compact kernel of characteristic length $L$ (Lesar et al., 2004, Rickman et al., 2006, Valdenaire et al., 2016). These studies seem to follow a suggestion by Groma on how to interpret the smooth dislocation density field (Groma et al., 1999). The most recent of these studies, (Valdenaire et al., 2016), has found that the spatial correlations in a dislocation system is dependent on the convolution length $L$. We will follow this formalism as well, with our approach most closely following that of Valdenaire et al. (Valdenaire et al., 2016), albeit at a significantly smaller length scale than they consider.

One of the major purposes behind this length scale dependent scheme is that it allows us to distinguish between dislocation structures which occur at two different scales. The local structure—at a length scale on the order of or below the convolution length $L$—can be associated with the correlation, while the spatial variation of the mean-field density can be used to describe longer-length structures such as dislocation patterning. Such patterns have been observed in some of the mean-field theories already presented (Groma et al., 1999, Xia et al., 2015a, Xia et al., 2015b). There is also evidence of these patterns in discrete dislocation dynamics simulations[1] (Deng et al., 2007). One of the major goals of the vector density continuum dislocation dynamics is to observe the formation of these patterns, as they are thought to play a significant role in the response of crystalline materials to monotonic and cyclic loading (Li et al., 2017, 2011, Sauzay et al., 2011). If these patterns can be reproducible, vector density continuum dislocation dynamics can be used in conjunction with its description of the finite deformation of crystals (Starkey et al., 2020) to solve micron scale plasticity problems such as crack initiation (Bao-Tong et al., 1989).

None of the above mean-field approaches are capable of capturing the true kinetics of a dislocation system. This is due to one unavoidable fact: dislocation interactions depend on the relative arrangement of dislocations, while mean-field theories all involve a systematic "forgetting" of this precise relative arrangement. This lost information regarding the relative arrangement of the dislocations can be represented by means of dislocation correlation functions, and is precisely the information which our present formulation purports to recover. The reason we wish to recover this information is that the kinetics of the dislocation system are strongly dependent on it through the energy functional of dislocation interactions. In mean-field approaches, the interaction energy and, correspondingly, the short-range stress field contain errors. Depending on the mean-field formulation chosen, this can represent different information that is lost. In the 2D density formulation, the density field is considered to vary only over distances significantly greater than the average dislocation spacing $1/\sqrt{\rho_0}$ (where $\rho_0$ is the average dislocation density of the entire crystal). In such cases, the short-range information the correlation recovers is the interaction of distinct dislocations. If a 3D, density-based approach is used and the density is allowed to vary faster than the average dislocation spacing, the mean-field still loses some information regarding connectivity and line tension effects. This has been seen to cause errors in the short-range stress field (Bertin, 2019). However, these missing elastic effects can be recovered from the mean field dislocation density and its gradients with certain integral moments of the correlation functions (Zaiser, 2015). As a result, there has been significant interest in calculating the form of the correlation functions.

[1] The authors of (Deng et al., 2007) use the term "pair-correlation" as a measure of these large length scale patterns. These are different statistics than we consider in the present work. For a closer analog of Deng et al.'s statistics, cf. (Groma et al., 1999) or (Csikor et al., 2008).

Two means have been explored to evaluate the correlation functions. The first follows statistical mechanical arguments to arrive at analytical forms of the dislocation-dislocation correlation function, while the other calculates the correlations brute-force from discrete simulations. Investigations along these two lines have elucidated some of the alterations which the correlation functions introduce into the dynamics in homogenous systems. In summary, the correlations produce additional stress terms (a friction and back stress in homogenous 2D systems (Groma et al., 2003, Valdenaire et al., 2016, Zaiser, 2015)), and alter the mobility of the mean-field density (Kooiman et al., 2015).

As mentioned, there have been attempts to analytically compute the geometrically necessary dislocation field induced in a homogenous dislocation field due to a dislocation pinned at the origin, controversially interpreted as a correlation. The analytical solutions obtained, however, still require a parameter which must be fit to discrete simulations (Groma et al., 2006, Limkumnerd et al., 2008, Zaiser, 2015) –for the clearest presentation of this parameter, see (Zaiser, 2015). As a result, one goal of the current work is to present a formalism by which these correlation functions might be computed directly from discrete dislocation configurations.

The question then arises as to how one might compute these correlation functions from discrete data. There have been several attempts to accomplish this task. They all involve the simulation of a random, homogenous distribution of discrete (2D) edge dislocations which have been relaxed at zero stress. The resulting relative separation vectors of same-sign and different-sign dislocations are binned into a histogram and normalized by the total dislocation content of the simulations. This is then as the correlation function. The first investigations which used this method (Gulluoglu et al., 1988, Wang et al., 1997) were largely motivated by a characterization of the dislocation microstructure, and agree with later evaluations of the correlation which arose with interest to the dynamics (Groma et al., 2006, Groma et al., 2003, Zaiser et al., 2001). However, the only attempt in 3-dimensions attempted to evaluate a radial distribution function of the scalar line density (Csikor et al., 2008), but it does not enter into the interaction energy calculation in a direct way. Valdenaire et al. (Valdenaire et al., 2016), using their convolution length dependent mean-field theory, were able to ascertain a dependence of the correlation on the convolution length using this binning method. A dependence on convolution length should be anticipated, as the convolution length controls the partition of relative arrangement information between the correlation and the density field: as the convolution length decreases, the mean-field density represents a better picture of the relative arrangement of the dislocations, and less correction is needed from the correlation functions. However, it also follows that by adjusting this convolution length, one may be observing qualitatively different relative arrangement information.

The present work represents an application of a convolution formalism approach to the high-resolution vector density theory of 3D dislocation arrangements, while in the process deepening the statistical underpinnings of the theory itself. The work may be outlined as follows: in section 2, we define a measure theoretic picture of the dislocation ensemble and the various densities, two-point distributions, and finally correlation functions which it produces; in section 3, we outline a means of evaluating the result from discrete simulations. In following sections, we apply this formalism to discrete dislocation configurations and present the correlation functions for dislocation pairs on like slip systems.

## Measure Theoretic Definition of Correlations

In order to arrive at a definition of the correlation function, we first motivate the discussion with a definition of the energy of a discrete dislocation configuration. We then follow with a discussion of the ensemble, mesoscopic density fields (mean-fields), and arrive at a definition of the correlation function which reveals a clear path forward in evaluation.

**Energy functional of a discrete dislocation configuration**

Let us consider a dislocated FCC crystal. The dislocation configuration represents 12 1-dimensional manifolds $\mathcal{L}^{[\alpha]}$ embedded in the crystal manifold $\mathcal{M}$, which we consider identical to $\mathbb{R}^3$. These manifolds represent the dislocations on each slip system $[\alpha]$. The elastic energy functional of the system $E$ can be expressed in terms of a double line integral over $\mathcal{L} = \bigcup_{[\alpha]} \mathcal{L}^{[\alpha]}$:

$$E = \frac{1}{2} \sum_{\alpha,\beta=1}^{12} \int_{\mathcal{L}^{[\alpha]}} dl \int_{\mathcal{L}^{[\beta]}} dl' \, \left( \boldsymbol{\xi}^{[\alpha]}(\boldsymbol{r}_l) \otimes \boldsymbol{\xi}^{[\beta]}(\boldsymbol{r}_{l'}) \right) : \boldsymbol{\mathcal{E}}^{[\alpha,\beta]}(\boldsymbol{r}_l - \boldsymbol{r}_{l'}) \, , \tag{1}$$

where $\boldsymbol{\xi}^{[\alpha]}$ denotes the unit tangent vector of $\mathcal{L}^{[\alpha]}$, and $\boldsymbol{\mathcal{E}}^{[\alpha,\beta]}$ denotes an energetic interaction kernel, a second rank tensor representing the energetic interaction of two differential segments $dl$ and $dl'$ on slip systems $[\alpha]$ and $[\beta]$, respectively. The interaction kernel is of the form (Hirth et al., 1982, Zaiser, 2015):

$$\boldsymbol{\mathcal{E}}^{[\alpha,\beta]}(\boldsymbol{\Delta r}) = \frac{\mu}{4\pi} \left\{ \left[ 2\boldsymbol{b}^{[\beta]} \otimes \boldsymbol{b}^{[\alpha]} - \boldsymbol{b}^{[\alpha]} \otimes \boldsymbol{b}^{[\beta]} \right] * \frac{1}{|\boldsymbol{\Delta}\mathbf{r}|} + \frac{1}{1-\nu} \left[ \left( \boldsymbol{b}^{[\alpha]} \times \boldsymbol{\nabla} \right) \otimes \left( \boldsymbol{b}^{[\beta]} \times \boldsymbol{\nabla} \right) \right] |\boldsymbol{\Delta r}| \right\} \tag{2}$$

Now stated, we will decline to use this expression in further analysis. For the sake of brevity, the dependence of equation (23) on the slip systems will be put aside to be reinserted at a later point in the analysis.

We choose to represent our system with a spatial field describing the density of lines around a given point in space (time dependence is implicit throughout the formulation presented here). As a result, we must define our basic (discrete) system in terms of a singular dislocation density $\varrho_i(\boldsymbol{r})$ which we will refer to as the discrete dislocation density:

$$\boldsymbol{\varrho}(\boldsymbol{r}) \coloneqq \int_{\boldsymbol{\mathcal{L}}} d\boldsymbol{l} \, \delta(\boldsymbol{r} - \boldsymbol{r}_l) \, , \tag{3}$$

where we have used the vector-valued differential line element $d\boldsymbol{l} \coloneqq dl \, \boldsymbol{\xi}(\boldsymbol{r}_l)$.

This dislocation density defines two measures on $\mathcal{M}$:

$$\mu_{\boldsymbol{\varrho}}(\Omega \subseteq \mathcal{M}) \coloneqq \int_{\Omega} \boldsymbol{\varrho}\, d^3\boldsymbol{r} = \int_{\mathcal{L}\cap\Omega} d\boldsymbol{l}, \tag{4}$$

$$\mu_{\varrho}(\Omega \subseteq \mathcal{M}) \coloneqq \int_{\Omega} |\boldsymbol{\varrho}|\, d^3\boldsymbol{r} = \int_{\mathcal{L}\cap\Omega} dl\,. \tag{5}$$

These measures represent the geometrically necessary dislocation content and total dislocation line length contained in $\Omega$, respectively. These are singular measures with respect to the volume measure, as they are non-zero on sets of zero volume (subsets of $\mathcal{L}$).

The density above allows us to re-express the energy functional by the following integration:

$$E = \frac{1}{2}\iint_{\mathcal{M}\times\mathcal{M}} \mathcal{E}_{ij}(\boldsymbol{r}-\boldsymbol{r}')\, \varrho_i(\boldsymbol{r})\varrho_j(\boldsymbol{r}')\, d^3\boldsymbol{r}'\, d^3\boldsymbol{r}\,, \tag{6}$$

where $i, j$ represent the vector components of $\boldsymbol{\varrho}(\boldsymbol{r})$ and Einstein's summation convention has is implied. In this form, it becomes apparent that the energy functional represents a sum of nine integrations of $\mathcal{E}_{ij}$ against nine measures $d\mu_{\varrho_i\varrho_j}$. These measures, however, are distinct from the measures in equations (5). Rather, $d\mu_{\varrho_i\varrho_j}$ represent measures of the product space $\mathcal{M}^2$. In the discrete case which we are considering, this product measure is simply expressed as the product of the discrete measures: $d\mu_{\varrho_i\varrho_j} = \varrho_i\varrho_j\, d^3\boldsymbol{r}\, d^3\boldsymbol{r}'$. However, we are interested in a statistical description of the dislocation configuration; in such a description, this product measure no longer has such a trivial form. In the following subsections, we will consider a definition of our statistical description, in the course of which it will be apparent why this product measure requires additional considerations. We will then return to examine equation (6) in light of this statistical description.

### A probabilistic definition of ensembles

While the "ensemble average" has been repeatedly employed in the discussion of dislocation dynamics, there has yet to be any rigorous definition of such an ensemble[2]. While this work may not completely arrive at such a lofty goal, it is the hope of the present authors that the following discussion will help to clarify what sort of entity this ensemble is.

[2] Interestingly, many previous treatments of continuum dislocation dynamics simply require an ensemble average with the linearity inherent to any projection operator. Namely, commutativity with the gradient operator.

The fundamental problem that we face in continuum dislocation dynamics is the fundamental problem of statistical mechanic, namely that of coarse-graining: given some limited information about a dynamical system (having thrown away more a detailed description), what conclusions can we draw from that limited information? Given an "ensemble" of microstates, what conclusions can we extract from the macrostate, some common property of this ensemble? In equilibrium considerations, this ensemble of microstates consists of copies of the system which are in some way equivalently prepared (Valdenaire et al., 2016, Zaiser, 2015). However, when we move into non-equilibrium considerations[3], this is no longer a useful analogy. When the average properties of the ensemble are changing, this implies change in the underlying microstates.

To elucidate the meaning of the non-equilibrium ensemble, we have to switch modes from thinking about equivalently prepared systems to conditioned probability spaces. An ensemble in the mathematical sense is a probability space conditioned by the level sets of a macrostate function. The reason that we have motivated this discussion by the above treatment of the discrete energy functional (equations 1 and 6) is that it informs the choice of macrostate variable: the dynamics of the coarse-grained variable are recoverable deterministically only if the energy is expressible in terms of the macrostate variable (Öttinger, 2005).

To be precise, an ensemble consists of four objects: 1) a space $\Gamma$ where the discrete arrangements are fully described; 2) a macrostate function $\Psi : \Gamma \to \mathbf{T}$ which represents a map from the microstate space to a (generally) lower-dimensional coarse-grained space T; 3) collections of subsets of the microstate space $\sigma(\Gamma)$ (these are technically $\sigma$-algebras, the details of which are in Appendix 1 and (Durrett, 2019)); and lastly, 4) a probability measure $P_\Psi$ which somehow uses the macrostate map to assign probabilities to all sets in $\sigma(\Gamma)$:

$$P_\Psi : \sigma(\Gamma) \to [0,1] \tag{7}$$

such that $P_\Psi(\Gamma)$ is equal to unity. This tuple, $(\Gamma, \Psi, \sigma(\Gamma), P_\Psi)$ is a sufficiently precise definition of what is meant by the ensemble. Not only can it express the statistical mechanics involved in equilibrium systems where $\Psi$ is time invariant, but also holds in the case of non-equilibrium systems where $\Psi$ has a non-trivial time evolution. The precision of this measure-theoretic definition of the ensemble is necessary to discuss the difficulties of the dislocation ensemble.

[3] By non-equilibrium we do not here mean thermal equilibrium (dislocations are never in any sort of thermal equilibrium), but rather the case in which the macrostate variables evolve with time.

As a probability space, the ensemble comes equipped with a projection operation which we will refer to as the ensemble average. The ensemble average represents a projection from functions of microstate variables $\gamma \in \Gamma$ to functions of macrostate variables $\psi \in \mathrm{T}$ in the following manner:

$$\langle A(\gamma)\rangle(\psi) \coloneqq \int_{\Gamma} dP_{\Psi}\, A(\gamma)\,. \tag{8}$$

There are two ways to define an ensemble in the traditional sense. In the first, the behavior of the microstates is analyzed on a single level set (e.g. the microcanonical ensemble where $\Psi \coloneqq E$). In the second, the ensemble average of $\Psi$ is constrained to a given form, and a probability distribution is chosen from the many candidates by means of another principle (e.g. the canonical ensemble, where $\Psi \coloneqq E$ and $\langle E\rangle \coloneqq \frac{1}{2}kT$). For a more detailed explanation of such ensembles, referred to as generalized microcanonical and generalized canonical ensembles, respectively, see (Öttinger, 2005).

We realize that the set theoretic notation used above may not be accessible to the average reader and as such it has been explained in Appendix 1 by demonstrating how this operation produces the microcanonical and Gibbs' canonical ensembles.

### A means of constructing a dislocation ensemble

As we have seen above, to construct a dislocation ensemble, we must first consider the space $\Gamma$ in which all possible discrete dislocation configurations are contained. We will then choose a macrostate function and constrain its ensemble average to arrive at some intuitions regarding the ensemble itself.

In the case of a dislocation configuration, a completely determined description is a state such as we have already discussed: the collection of the twelve line-objects corresponding to the twelve species of dislocation line $\left\{\mathcal{L}^{[\alpha]}\right\}_{\alpha=1}^{12}$. The space of microstates, then, is the set of all space curves, with a few minor constraints regarding being confined to the slip planes as well as the non-termination requirement.

Now let us choose a macrostate function. Instead of a scalar functional like the total line content, the plastic strain, or the energy of a simulation box, we choose the discrete dislocation density distribution $\boldsymbol{\varrho}(\boldsymbol{r})$. This represents a map from the space of dislocation configurations to the space of vector-valued distributions in $\mathbb{R}^3$:

$$\Psi: \{\mathcal{L}\} \to D^3_{\mathbb{R}^3}, \Psi(\mathcal{L}) \coloneqq \boldsymbol{\varrho}_{\boldsymbol{\mathcal{L}}}(\boldsymbol{r}) \tag{9}$$

In a similar operation by which the canonical ensemble is constructed by enforcing a certain average value of the energy, we may construct our dislocation ensemble by constraining two ensemble averages of $\boldsymbol{\varrho}_{\mathcal{L}}(\boldsymbol{r})$ to a particular distribution, which we will call the mean-field dislocation density vector $\boldsymbol{\rho}(\boldsymbol{r})$. The reader is advised to distinguish the calligraphic $\boldsymbol{\varrho}$—the discrete density field—and $\boldsymbol{\rho}$—the coarse-grained density field. Such a distinction is significant to the remainder of the work. The ensemble is then defined by two constraints. Firstly, we constrain the vector ensemble average, and secondly, we constrain the magnitude average by introducing a probability distribution $P_{\mathrm{LBE}}$ such that:

$$\langle \boldsymbol{\varrho}_{\boldsymbol{\mathcal{L}}}(\boldsymbol{r}) \rangle = \int P_{\mathrm{LBE}}(d\mathcal{L}) \boldsymbol{\varrho}_{\boldsymbol{\mathcal{L}}}(\boldsymbol{r}) \coloneqq \boldsymbol{\rho}(\boldsymbol{r}), \tag{10}$$

$$\left\langle |\hat{\boldsymbol{a}} \cdot \boldsymbol{\varrho}_{\mathcal{L}}(\boldsymbol{r})| \right\rangle = \int P_{\mathrm{LBE}}(d\mathcal{L}) |\hat{\boldsymbol{a}} \cdot \boldsymbol{\varrho}_{\boldsymbol{\mathcal{L}}}(\boldsymbol{r})| \coloneqq |\hat{\boldsymbol{a}} \cdot \boldsymbol{\rho}(\boldsymbol{r})| \tag{11}$$

where $\hat{\boldsymbol{a}}$ represents any constant vector. We refer to this second constraint as the "line bundle" constraint, as it implies that all the microstate densities are roughly parallel to the mesoscopic density vector. This constraint eliminates the possibility of so-called "statistically stored dislocations" by disallowing geometric cancellation of dislocation densities in the ensemble average. Generally, formalisms have been used where the total scalar dislocation density at a point is not necessarily equal to the magnitude of the vector density (Hochrainer, 2015, Zaiser, 2015). This means that we are treating a different ensemble than these formalisms; our goal in the present work is not to present an ensemble consistent with such formalisms, only with our own. However, these requirements do impose a significant constraint on the choice of the vector density field, namely the spatial scale on which it is allowed to vary (Lin et al., 2020, Xia et al., 2015).

In the line bundle constraint, there was no mention of a specific choice of mean-field density. Any smooth density field may be chosen as long as it meets some criteria which are necessary for it to be consistent with the underlying dislocation objects. Firstly, it must be solenoidal (i.e. $\nabla \cdot \rho = 0$) due to the requirement that dislocations not terminate within the crystal; this condition may be somewhat relaxed if one desires to treat dislocation networks. Secondly, the density must be able to vary in space on a scale at which dipoles annihilate (taken to be on the order of 50-150 nm).

There exists an operation by which a density field which meets these criteria *a priori* can be created. Begin with some parent dislocation configuration $\mathcal{L}_0$ with discrete density $\boldsymbol{\varrho}_0(\boldsymbol{r})$. Define the vector density field as a convolution of this parent configuration with a weight function $w_L(\boldsymbol{r})$ with compact support $\Omega_L$ characterized by some length parameter $L$ which we will refer to as the convolution length. The vector density field is then defined as:

$$\boldsymbol{\rho}(\boldsymbol{r}) \coloneqq (w_L * \boldsymbol{\varrho_0})(\boldsymbol{r}) = \int_{\Omega_L} w_L(\boldsymbol{r}')\boldsymbol{\varrho}_0(\boldsymbol{r} - \boldsymbol{r}')\, d^3\boldsymbol{r}' \,. \tag{12}$$

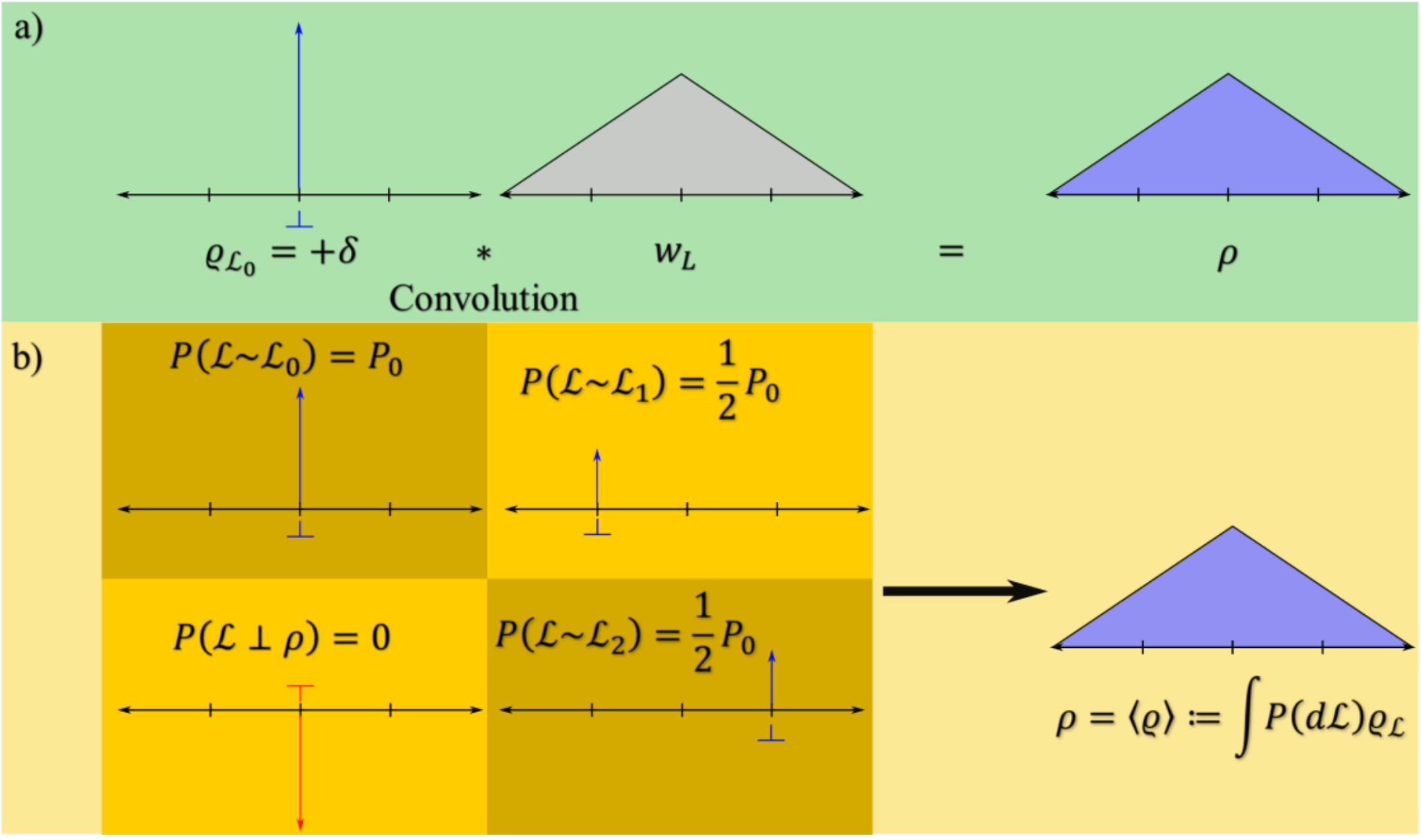

**Fig. 1** An overview of the formation of the line bundle ensemble by a generalized canonical approach. In part a), note that a single parent microstate $\mathcal{L}_0$ is used to generate the mean-field density $\boldsymbol{\rho}(\boldsymbol{r})$ by convolution with some weight function $w_L$, i.e. $\boldsymbol{\rho}(\boldsymbol{r}) := (\boldsymbol{\varrho} * w_L)(\boldsymbol{r})$. In part b), we show a toy model of the induced probability measure on the space of dislocation configurations. The ensemble average of the discrete density (i.e. integration against this probability measure in the space of line configurations), is constrained to equal the mean field density. By equating the ensemble average to the field generated by means of the convolution operation, we can induce a probability measure by means of some maximum entropy argument. One property of the probability measure, however, is guaranteed by the line bundle constraint: the class of states $\{\mathcal{L} \perp \boldsymbol{\rho}\}$ (for which the discrete density is of the opposite sign as $\boldsymbol{\rho}$), is necessarily of null probability (see Appendix 2).

The solenoidality of the dislocation density is inherited from the discrete density distribution. However, in accordance with the variation length considerations mentioned above, the length $L$ is considered to be on the order of 50-150 nm. We would like to point out that the use of the convolution operation does not replace our statistical discussion of probability measures and ensemble averages. Rather, by constraining the ensemble average to this convoluted density, the probability measure is induced (cf. Appendix 1 where a similar operation is performed to obtain Gibbs' canonical ensemble). For a brief recapitulation of this process, consult Fig. 1.

We further note that this small convolution length is in large part where the present work (as well as previous works by the second author) may differ from coarser resolution models. In previous works explicitly considering a convolution of the discrete configuration, $L$ was chosen to be on the order of the mean dislocation spacing in the simulation volume (Valdenaire et al., 2016). enabled those authors to apply equilibrium arguments like local homogeneity and steady state flow. However, the purpose of the present model is essentially to solve for the transient flows in a microscopic continuum context, for reasons already outlined in the introduction. Moreover, longer convolution lengths would need to be treated with a model which allows multiple dislocation line directions at a single point, e.g. (Hochrainer, 2015).

**Product measures and correlation**

The previous sections served to elucidate the statistical constructs implicit in defining a vector density field. We now wish to return to the problem of the recoverability of the energy (and thereby the kinetics) from the mesoscopic dislocation density vector. This requires an examination of $\langle E \rangle$. By linearity of the integral, this is simply:

$$\langle E \rangle = \frac{1}{2} \iint_{\mathcal{M}\times\mathcal{M}} \mathcal{E}_{ij}(\boldsymbol{r}-\boldsymbol{r}') \left\langle \varrho_i(\boldsymbol{r}) \varrho_j(\boldsymbol{r}') \right\rangle d^3\boldsymbol{r}'\, d^3\boldsymbol{r}\,. \tag{13}$$

Let us examine this expression. This represents a sum of nine integrations against 9 measures (or 4 if the dislocation densities are planar) on $\mathcal{M} \times \mathcal{M}$. We make the assumption that these nine integrations are each recoverable rather than only being preserved in sum. As such, we will treat them individually, and denote the treatment of an individual component by replacing the indices with an asterisk:

$$\langle E_* \rangle = \frac{1}{2} \iint_{\mathcal{M}\times\mathcal{M}} \mathcal{E}_*(\boldsymbol{r}-\boldsymbol{r}') \left\langle \varrho_*(\boldsymbol{r}) \varrho_*(\boldsymbol{r}') \right\rangle d^3\boldsymbol{r}'\, d^3\boldsymbol{r}, \quad \langle E \rangle = \sum_{*=1}^{9} \langle E_* \rangle\,. \tag{14}$$

Now we are interested in the measure produced by the two-point density, i.e. $d\nu_{\text{prod}} = \langle \varrho_*(\boldsymbol{r}) \varrho_*(\boldsymbol{r}') \rangle\, d^3\boldsymbol{r}'\, d^3\boldsymbol{r}$. We would like, however, to integrate against the "naïve" two-point density measure with similar indices, i.e. $d\mu_{\text{naïve}} = \rho_*(\boldsymbol{r}) \rho_*(\boldsymbol{r}')\, d^3\boldsymbol{r}'\, d^3\boldsymbol{r}$. To perform such a transformation, we must first examine why these two are not equal. In the process, we will understand precisely what information was lost in the averaging process.

A line $\mathcal{L}$ contains not just information regarding the configuration at a single point, but at all tuples of points. If two points lie on the line, the tuple will be contained in $\mathcal{L} \times \mathcal{L}$. Similarly, if a set of *n* points $(\boldsymbol{r}_1, \boldsymbol{r}_2, \ldots\, \boldsymbol{r}_n)$ all fall on the line, this tuple will be contained in $\mathcal{L}^n$. This property also holds for the discrete density associated with $\mathcal{L}$; the multi-point distribution $\boldsymbol{\varrho}(\boldsymbol{r}_1) \otimes \boldsymbol{\varrho}(\boldsymbol{r}_2) \otimes \ldots \boldsymbol{\varrho}(\boldsymbol{r}_n) d^3\boldsymbol{r}_1 d^3\boldsymbol{r}_2 \ldots d^3\boldsymbol{r}_n$ contains the information regarding the configuration at all these points simultaneously. However, when defining the ensemble average, we did not constrain the multi-point distributions, only the single-point distribution $\langle \boldsymbol{\varrho}(\boldsymbol{r}) \rangle$.

To see why we lost this multi-point information in the course of ensemble averaging, examine the ensemble average which results in the two-point density:

$$\left\langle \boldsymbol{\varrho}(\boldsymbol{r}) \otimes \boldsymbol{\varrho}(\boldsymbol{r}') \right\rangle = \int P_{\mathrm{LBE}}(d\mathcal{L}) \big( \boldsymbol{\varrho}_{\mathcal{L}}(\boldsymbol{r}) \otimes \boldsymbol{\varrho}_{\mathcal{L}}(\boldsymbol{r}') \big). \quad (15)$$

We may express this as an iterated integral by means of the identity convolution operation with respect to $P_{\mathrm{LBE}}$.

$$\left\langle \boldsymbol{\varrho}(\boldsymbol{r}) \otimes \boldsymbol{\varrho}(\boldsymbol{r}') \right\rangle = \int P_{\mathrm{LBE}}(d\mathcal{L}) \int P_{\mathrm{LBE}}(d\mathcal{L}') \delta(\mathcal{L} - \mathcal{L}') \left( \boldsymbol{\varrho}_{\mathcal{L}}(\boldsymbol{r}) \otimes \boldsymbol{\varrho}_{\mathcal{L}'}(\boldsymbol{r}') \right). \quad (16)$$

This Dirac delta object $\delta(\mathcal{L} - \mathcal{L}')$ ensures that the two-point density examines the interaction between densities at two points only if those densities belong to the same microstate. The naïve two-point density, however, does not contain this nuance:

$$\begin{aligned} \boldsymbol{\rho}(\boldsymbol{r}) \otimes \boldsymbol{\rho}(\boldsymbol{r}') &= \langle \boldsymbol{\varrho}(\boldsymbol{r}) \rangle \otimes \langle \boldsymbol{\varrho}(\boldsymbol{r}') \rangle \\ &= \left( \int P_{\mathrm{LBE}}(d\mathcal{L}) \boldsymbol{\varrho}_{\mathcal{L}}(\boldsymbol{r}) \right) \otimes \left( \int P_{\mathrm{LBE}}(d\mathcal{L}') \boldsymbol{\varrho}_{\mathcal{L}'}(\boldsymbol{r}') \right) \end{aligned}$$

$$\boldsymbol{\rho}(\boldsymbol{r}) \otimes \boldsymbol{\rho}(\boldsymbol{r}') = \int P_{\mathrm{LBE}}(d\mathcal{L}) \int P_{\mathrm{LBE}}(d\mathcal{L}') \left( \boldsymbol{\varrho}_{\mathcal{L}}(\boldsymbol{r}) \otimes \boldsymbol{\varrho}_{\mathcal{L}'}(\boldsymbol{r}') \right). \quad (17)$$

As a result, the naïve two-point density contains information regarding interaction across microstates. This is obviously unphysical, as the energy of a dislocation configuration cannot be dependent on these ghostly interactions with other possible configurations.

However, this is not to say that we are unable to capture the true interactions by integrating against the naïve product measure; we may do so in the following way:

$$\nu_{\mathrm{prod}}(A) = \int_A d\nu_{prod} = \int_A g \, d\mu_{\mathrm{naïve}} \ \forall A \subseteq \mathcal{M}, \quad (18)$$

where $g(\boldsymbol{r}, \boldsymbol{r}')$ here denotes the two-point correlation, which is a Radon-Nikodym derivative. There is a condition on the existence of this function, which is that $\nu_{\mathrm{prod}}$ must be "absolutely continuous" with respect to $\mu_{\mathrm{naïve}}$ (Durrett, 2019). That is, for all $A \subseteq \mathcal{M}$ such that $\mu_{\mathrm{naïve}}(A) =$ 0, $\nu_{\text{prod}}(A)$ must also equal zero. This is guaranteed as a corollary of the line bundle constraint. See Appendix 2 for a rigorous derivation of this property of line bundle ensembles.

For our case in which both $\nu_{\text{prod}}$ and $\mu_{\text{naïve}}$ have density functions, the Radon-Nikodym derivative is simply expressible as:

$$g^{(*,*)}(\boldsymbol{r},\boldsymbol{r}') = \frac{d\nu_{\text{prod}}}{d\mu_{\text{naïve}}} = \frac{\langle \varrho_*(\boldsymbol{r})\varrho_*(\boldsymbol{r}')\rangle}{\rho_*(\boldsymbol{r})\rho_*(\boldsymbol{r}')}. \tag{19}$$

This function is protected from singularity in the denominator precisely by the absolute continuity property. It represents in some sense a spatial correction for the erroneous ghost interactions between microstates. It does this by applying a scale factor to the naïve two-point density, i.e. at some pairs of points $\boldsymbol{r},\boldsymbol{r}'$ the naïve two-point density is likely to be due to the ghost interactions: at such locations $g(\boldsymbol{r},\boldsymbol{r}')$ will be less than unity. At other pairs of points, the naïve two-point density might underpredict the true interactions: at such locations $g(\boldsymbol{r},\boldsymbol{r}')$ will be greater than unity.

We further note that the 'mean-field' densities which appear in the denominator of equation (18) are in fact local fields. Like the introduction of the line bundle constraint, this is a point of departure from treatments which consider the denominator to be the average density over an infinite domain (cf. (Deng et al., 2007, Stoyan et al., 1986, Zaiser et al., 2001) ). The treatment in (Valdenaire et al., 2016) does use a local mean-field density, but at a significantly different resolution.

This definition does not involve any form of tensor summation. Some may object to the use of vector index notation in the above equation, but we have considered each component of $\rho_i(\boldsymbol{r})\rho_j(\boldsymbol{r}')$ and $\varrho_i(\boldsymbol{r})\varrho_j(\boldsymbol{r}')$ as separate scalar quantities throughout this discussion of measures. This equation is no exception to that rule. The above equation is best understood in the sense that $\langle \varrho_i(\boldsymbol{r})\varrho_j(\boldsymbol{r}')\rangle = g^{(i,j)}(\boldsymbol{r},\boldsymbol{r}')\rho_i(\boldsymbol{r})\rho_j(\boldsymbol{r}')$, with $g^{(i,j)}$ being a scalar correlation transforming the individual components of the tensor product $\rho_i(\boldsymbol{r})\rho_j(\boldsymbol{r}')$ of the mean field densities to the equivalent components of the two-point density, which is also a tensor. We also note that the ensemble average is a linear operator, since it is simply integration against a probability measure on the microstates (cf. equations (8,10,11)). Since, however, the mean-field product by definition is a constant function on the microstate space, it commutes with the ensemble average resulting in the following relation:

$$g^{(i,j)}(\boldsymbol{r},\boldsymbol{r}') = \left\langle \frac{\varrho_i(\boldsymbol{r})\varrho_j(\boldsymbol{r}')}{\rho_i(\boldsymbol{r})\rho_j(\boldsymbol{r}')} \right\rangle = \langle \tilde{\varrho}_i(\boldsymbol{r})\tilde{\varrho}_{\boldsymbol{j}}(\boldsymbol{r}')\rangle, \tag{20}$$

where we have introduced a field which we will refer to as a "protocorrelation density"

$$\tilde{\varrho}_i(\boldsymbol{r}) \coloneqq \frac{\varrho_i(\boldsymbol{r})}{\rho_i(\boldsymbol{r})}. \tag{21}$$

Thus, we have arrived at a form of the correlation which informs us how to evaluate it from discrete data. We notice that the protocorrelation is a non-negative function, as the sign of $\varrho_i(\boldsymbol{r})$ and $\rho_i(\boldsymbol{r})$ are identical as a corollary of the line bundle constraint. Given some way of evaluating this ensemble average, we must merely examine the average product of protocorrelation densities at two points.

The energy of the system (integration of the interaction kernel against the product measure), can now be expressed as:

$$\langle E\rangle = \frac{1}{2}\sum_{ij}\iint_{\mathcal{M}\times\mathcal{M}} \mathcal{E}_{ij}(\boldsymbol{r}-\boldsymbol{r}')\, g^{(i,j)}(\boldsymbol{r},\boldsymbol{r}')\rho_i(\boldsymbol{r})\rho_j(\boldsymbol{r}')\, d^3\boldsymbol{r}'\, d^3\boldsymbol{r}\,. \tag{22}$$

Before we move on, however, we reintroduce in a straightforward manner the multi-slip aspect of the dislocation configuration (previously dropped from equation (1)) in the following two equations:

$$\langle E\rangle = \frac{1}{2}\sum_{\alpha,\beta=1}^{12}\sum_{ij}\iint_{\mathcal{M}\times\mathcal{M}} \mathcal{E}_{ij}^{[\alpha,\beta]}(\boldsymbol{r}-\boldsymbol{r}')\, g^{(i,j)[\alpha,\beta]}(\boldsymbol{r},\boldsymbol{r}')\rho_i^{[\alpha]}(\boldsymbol{r})\rho_j^{[\beta]}(\boldsymbol{r}')\, d^3\boldsymbol{r}'\, d^3\boldsymbol{r}\,, \tag{23}$$

with

$$g^{(i,j)[\alpha,\beta]}(\boldsymbol{r},\boldsymbol{r}') = \left\langle \tilde{\varrho}_i^{[\alpha]}(\boldsymbol{r})\tilde{\varrho}_j^{[\beta]}(\boldsymbol{r}') \right\rangle. \tag{24}$$

## Evaluation Scheme

The evaluation of the ensemble average in equation (23) is not by any means trivial. The particular dislocation ensemble which we have constructed in the present work cannot be realized by the simple superposition of many simulation boxes of discrete dislocations onto one another. In fact, we only ever have access to a single microstate: the parent configuration in equation (12). In the present section we present a scheme by which we may evaluate the expression for the correlation function seen in equation (23). This is a two-step process. The first step involves a discretization scheme in which we mollify the singular densities present in $\tilde{\varrho}_i^{[\alpha]}(\boldsymbol{r})$**.** The second step is to define a certain statistical homogeneity assumption that will allow us to empirically measure the underlying random variable $\tilde{\varrho}_i^{[\alpha]}(\boldsymbol{r})\tilde{\varrho}_j^{[\beta]}(\boldsymbol{r}')$ using only the parent configuration.

### Regularization scheme

In order to evaluate any expression containing $\tilde{\varrho}_i^{[\alpha]}(\boldsymbol{r})$ from simulation data, we must mollify the singular character of the discrete density $\varrho_i^{[\alpha]}(\boldsymbol{r})$. In order to perform this, we perform a double convolution with some weight function $w_0$, suppressing for the moment the slip system notation:

$$w_0 * g^{(i,j)}(\boldsymbol{r},\boldsymbol{r}') * w_0' = \langle (\tilde{\varrho}_i * w_0)(\boldsymbol{r})(\tilde{\varrho}_i * w_0')(\boldsymbol{r}') \rangle. \tag{25}$$

The prime or lack thereof denotes whether the convolution is over $\boldsymbol{r}'$ or $\boldsymbol{r}$, respectively. The weight function is arbitrary, so long as it has unit integral and is of small, compact support characterized by some length $l_0$. If the convolution length in the mean field calculation ($L$) is significantly longer than $l_0$, then we may treat the mean field density as constant over the support of $w_0$, simplifying our expression of $\tilde{\varrho} * w_0$:

$$\tilde{\varrho}_i^* \coloneqq \tilde{\varrho}_i * w_0 \approx \frac{\varrho_i * w_0}{\rho_i}. \tag{26}$$

Note that we have incorporated the weight function convolution into a compact notation.

### Empirical measurement

To understand how we can empirically measure the correlation, let us examine how empirical measurements of a random variable are made. For clarity, we will proceed with several definitions which are quite standard. Given some random variable $X$ and $n$ independent measurements of that variable $X_i$, we may be confident that by the law of large numbers the empirical mean approaches the ensemble average:

$$\bar{X} \coloneqq \frac{\sum_{i=1}^{n} X_i}{n} \to \mu \coloneqq \langle X \rangle, \tag{27}$$

$$\overline{X^2} \coloneqq \frac{\sum_{i=1}^{n} (X_i - \mu)^2}{n} \to \varsigma^2 \coloneqq \mathrm{Var}(X). \tag{28}$$

We also know that by the central limit theorem, the following normalized sum converges in distribution to a standard normal random variable:

$$\frac{\sum_{i=1}^{n} (X_i - \mu)}{\varsigma\sqrt{n}} \to N(0,1). \tag{29}$$

It follows by continuity of the inverse square root that:

$$\frac{\sum_{i=1}^{n} (X_i - \mu)}{n} \frac{n}{\sqrt{\sum_{i=1}^{n} (X_i - \mu)^2}} \to N(0,1). \tag{30}$$

As a result, we may quantify our uncertainty by noting that 68% of measurements will fall in the range:

$$\bar{X} \pm \sigma, \tag{31}$$

where $\sigma \coloneqq \sqrt{\overline{X^2}}/n$ we define as the standard error of the empirical measurement.

If we examine the average we wish to compute, a path forward in identifying independent measurements should become apparent. Consider first the case where only one slip system is present in the crystal. In this case, the averages which we would like to consider are of the form:

$$g^{(i,j)}(\boldsymbol{r}, \boldsymbol{r}') = \left\langle \tilde{\varrho}_i^*(\boldsymbol{r}) \tilde{\varrho}_j^*(\boldsymbol{r}') \right\rangle. \tag{32}$$

If we can form some collection of independent measurements of this random variable $\left\{\tilde{\varrho}_i^*(\boldsymbol{r}) \tilde{\varrho}_j^*(\boldsymbol{r}')\right\}_{k=1}^{N}$, we can apply the law of large numbers to obtain the mean as:

$$\frac{1}{N} \sum_{k=1}^{N} \left[\tilde{\varrho}_i^*(\boldsymbol{r}) \tilde{\varrho}_j^*(\boldsymbol{r}')\right]_k \xrightarrow{N \to \infty} \left\langle \tilde{\varrho}_i^*(\boldsymbol{r}) \tilde{\varrho}_j^*(\boldsymbol{r}') \right\rangle \tag{33}$$

and even obtain a confidence interval of this calculation by means of the central limit theorem.

If the random variable $\tilde{\varrho}_i^*(\boldsymbol{r})\tilde{\varrho}_j^*(\boldsymbol{r}')$ is independent of the information local to any point in the crystal, we can treat spatial observations of the right-hand side of equation (32) as independent. The operation of defining the ensemble by using the density vector field as the macrostate constraint in another sense defines the density vector field to be the only available local information. As a result, it is sufficient to assume that the functional form of $g^{(i,j)}(\boldsymbol{r}, \boldsymbol{r} + \boldsymbol{\Delta r})$ is independent of the density field at $\boldsymbol{r}$ or $\boldsymbol{r}'$. In more rigorous terms, to treat spatial observations of $\tilde{\varrho}_i^*(\boldsymbol{r})\tilde{\varrho}_j^*(\boldsymbol{r} + \boldsymbol{\Delta r})$ as independent events, one must only assume that:

$$\frac{\delta g^{(i,j)}(\boldsymbol{r}, \boldsymbol{r} + \boldsymbol{\Delta r})}{\delta \rho(\boldsymbol{r})} = 0, \tag{34}$$

where the $\delta$ operator here represents the variational derivative. The variational derivative above amounts simply to the statement that by expressing the two-point density as $\langle \varrho_i(\boldsymbol{r})\varrho_j(\boldsymbol{r})\rangle = \rho_i(\boldsymbol{r})\rho_j(\boldsymbol{r}')g^{(i,j)}(\boldsymbol{r}, \boldsymbol{r}')$, we have explicitly considered the density dependence of the two-point density.

The implication of this assumption is that the correlation function describes the average neighborhood of a unit density component regardless of the magnitude of the density. We will refer to this assumption as a statistical homogeneity assumption, as it allows us to treat all points with non-zero density as equivalent measurements of the correlation product. Moreover, it allows us to drop the spatial dependence on $\boldsymbol{r}$ and to consider the correlation as only a function of the separation distance $\boldsymbol{\Delta r} \coloneqq \boldsymbol{r} - \boldsymbol{r}'$:

$$g^{(i,j)}(\boldsymbol{r}, \boldsymbol{r}') \to g^{(i,j)}(\boldsymbol{\Delta r}) \tag{35}$$

A correlation of this form represents a sort of average atmosphere of $j$-component density field surrounding a point of non-zero $i$-component density. As a result, it makes intuitive sense that we may observe this average atmosphere by examining a large number of local atmospheres surrounding a large number of points in a single microstate (namely, the parent microstate $\mathcal{L}_0$).

## Sample classes in the simulated crystal

At this point, we choose to make explicit what is meant by observing spatial points in the simulated crystal and how these are used to compute the correlation function using the methods discussed in the previous subsection. Choosing a sample grid $S$ to be a finite, countable collection of position vectors:

$$S \coloneqq \{\boldsymbol{r}_l\}_{l=1}^{N}, \tag{36}$$

we then have a measurement of $\tilde{\varrho}_i^*(\boldsymbol{r})\tilde{\varrho}_j^*(\boldsymbol{r'})$ at all points in $S \times S$, $N^2$ in total.

To create classes which contain observations of the same correlation average, we consider two factors. We consider points equivalent if they are kinetically or kinematically equivalent up to the value of the microscopic dislocation density. We call points kinetically equivalent if the interaction energy would be equivalent, and we call points kinematically equivalent if the transport relations of the discrete density would be equivalent at the two points.

We will now translate these requirements into partitions of our measurement space $S \times S$. Consider first the kinetic equivalence classes. Examining the dependency of the interaction kernel $\mathcal{E}_{ij}$ (equation (2)) allows us to note that two pairs of points are kinetically equivalent if they share the same separation vector $\boldsymbol{r} - \boldsymbol{r'}$. This allows us to partition $S \times S$ into the equivalence classes:

$$(S \times S)_{\boldsymbol{\Delta r}} \coloneqq \{(\boldsymbol{r}, \boldsymbol{r}') \in S \times S : \boldsymbol{r} - \boldsymbol{r}' = \boldsymbol{\Delta r}\}. \tag{37}$$

Furthermore, we note that the transport behavior of a location with zero density field is significantly different from locations where there is a density field present. Thus we consider the set:

$$\tilde{S} \coloneqq \{\boldsymbol{r} \in S : \ \varrho_i(\boldsymbol{r}) \neq 0 \ \text{ for } i = 1, 2, 3\}. \tag{38}$$

Considering the points which are kinetically *and* kinematically equivalent results in a partition of the measurement space into sets of interest:

$$\pi_{\boldsymbol{\Delta r}} \coloneqq \{(\boldsymbol{r}, \boldsymbol{r}') \in \tilde{S} \times \tilde{S} : \boldsymbol{r} - \boldsymbol{r}' = \boldsymbol{\Delta r}\} = (S \times S)_{\boldsymbol{\Delta r}} \cap (\tilde{S} \times \tilde{S}) \tag{39}$$

and irrelevant sets $(S \times S) \setminus (\tilde{S} \times \tilde{S})$.

Treating these sets of interest as equivalent measurements of $\tilde{\varrho}_i^*(\boldsymbol{r})\tilde{\varrho}_j^*(\boldsymbol{r}+\boldsymbol{\Delta r})$, we may now apply the law of large numbers and central limit theorem to the correlation expression in equation (32):

$$w_0 * g^{(i,j)}(\boldsymbol{\Delta r}) * w_0' = \lim_{n_{\boldsymbol{\Delta r}}\to\infty} \frac{1}{n_{\boldsymbol{\Delta r}}} \sum_{\pi_{\boldsymbol{\Delta r}}} \tilde{\varrho}_i^*(\boldsymbol{r}_{\pi_{\boldsymbol{\Delta r}}})\tilde{\varrho}_j^*(\boldsymbol{r}'_{\pi_{\boldsymbol{\Delta r}}}), \tag{40}$$

$$\sigma^{(i,j)}(\boldsymbol{\Delta r}) = \lim_{n_{\boldsymbol{\Delta r}}\to\infty} \frac{1}{n_{\boldsymbol{\Delta r}}} \left\{ \sum_{\pi_{\boldsymbol{\Delta r}}} \left[\tilde{\varrho}_i^*(\boldsymbol{r}_{\pi_{\boldsymbol{\Delta r}}})\tilde{\varrho}_j^*(\boldsymbol{r}'_{\pi_{\boldsymbol{\Delta r}}}) - w_0 * g^{(i,j)}(\boldsymbol{\Delta r}) * w_0'\right]^2 \right\}^{\frac{1}{2}}, \tag{41}$$

where we have represented by $n_{\boldsymbol{\Delta r}}$ the cardinality of $\pi_{\boldsymbol{\Delta r}}$. We note two things regarding the standard error functions. First, this standard error does not represent the standard error in the calculation of $g^{(i,j)}(\boldsymbol{\Delta r})$ but rather in the double convolution $w_0 * g^{(i,j)}(\boldsymbol{\Delta r}) * w_0'$. Secondly, the standard error varies with the components $i$ and $j$ as well as being a spatially varying field.

**Multi-slip considerations**

Further discrimination among our sample points becomes necessary when we consider systems with twelve slip system protocorrelations $\tilde{\varrho}_i^{[\alpha]}(\boldsymbol{r})$ (we momentarily suppress the convolution notation in favor of slip system dependence). We first consider an altered form of $\tilde{S}$ which is unique to each slip-system:

$$\tilde{S}^{[\alpha]} \coloneqq \left\{\boldsymbol{r} \in S :\ \varrho_i^{[\alpha]}(\boldsymbol{r}) \neq 0 \ \text{ for } 1 \leq i \leq 3\right\}. \tag{42}$$

Measurements of the product $\tilde{\varrho}_i^{[\alpha]}(\boldsymbol{r})\tilde{\varrho}_j^{[\beta]}(\boldsymbol{r})$ must therefore be elements of $\tilde{S}^{[\alpha]} \times \tilde{S}^{[\beta]}$.

An additional constraint on kinematically equivalent points also emerges only in the multi-slip case. Sessile dislocation segments, having a Burgers vector which is a sum of two of the slip system Burgers vectors $\boldsymbol{b}_{\text{sess}} = \boldsymbol{b}^{[\alpha]} \pm \boldsymbol{b}^{[\beta]}$. As such, such a segment can be represented by two overlapping densities at that point, i.e. $\left|\varrho_i^{[\alpha]}(\boldsymbol{r}_{\text{sess}})\right|, \left|\varrho_i^{[\beta]}(\boldsymbol{r}_{\text{sess}})\right| > 0$. However, in order to take into account their limited kinematics (lengthening along the intersection line between slip planes) the dislocation density transport equations must be suspended at these points. For this reason, we do not consider these points equivalent to points where only one density field is present.

To be precise in this omission, consider the sets $G^{[\alpha]}$ and $J^{[\alpha,\beta]}$ (glissile and junction):

$$J^{[\alpha.\beta]} = \tilde{S}^{[\alpha]} \cap \tilde{S}^{[\beta]} \text{ for } \alpha \neq \beta, \tag{43}$$

$$G^{[\alpha]} = \tilde{S}^{[\alpha]} \setminus \bigcup_{\alpha>\beta=1}^{12} J^{[\alpha,\beta]} \ . \tag{44}$$

Thus, there are three new types of pairs in our sample partition: glissile-glissile pairs, junction-junction pairs, and glissile-junction pairs. We will in the present work only consider the glissile-glissile pairs:

$$\pi_{\boldsymbol{\Delta r}}^{[\alpha,\beta]gg} \coloneqq \{(\boldsymbol{r}, \boldsymbol{r}') \in G^{[\alpha]} \times G^{[\beta]} : \boldsymbol{r} - \boldsymbol{r}' = \boldsymbol{\Delta r}\}. \tag{45}$$

To summarize, in a multi-slip dislocation system, there are several classes of correlation functions, calculated as in equation (40) with varying types of pair sets considered. Most broadly there are the glissile correlations and the junction correlations, the distinction of which we have treated immediately above. Secondly, there are what we will refer to as self-correlations and cross-correlations, considering like-slip-system densities and unlike-slip system densities, respectively. That is, their pair sets are of the form of equation (45) with $\beta = \alpha$ and $\beta \neq \alpha$.

## Calculations

As a preliminary consideration, we have calculated a small subset of the correlation functions from a set of discrete dislocation dynamics simulations. Specifically, we consider the glissile self-correlations only.

### Dislocation dynamics simulations

Discrete dislocation dynamics simulations of copper were carried out using microMegas (Devincre et al., 2011). A total of 45 simulations were performed, all beginning from initial configurations of dipolar loops in a periodic box of dimensions 4.4 × 4.90 × 5.8 μm. The dipolar loop configurations consist of four edge dislocations on two slip systems, all 1 μm in length. 15 distinct seed numbers were used in the pseudorandom number generator in order to create the initial configurations, and simulations were run in a strain-controlled mode to 0.3% plastic strain. Parameters used to create the initial configurations can be found in Table 1, while simulation parameters for the dislocation dynamics simulations can be found in Table 2. Each configuration was subjected to 3 simulations with tensile loading in the [100], [010], and [001] directions respectively. This was done to suppress any dependence which may have arisen in the correlations due to the loading direction, as such a dependence has been seen to occur in 2D dislocation dynamics simulations (Valdenaire et al., 2016). These 15 initial configurations were each subject to three simulations, resulting in 45 dislocation trajectories for analysis.

These 45 simulations do not represent 45 different microstates which we take to be representative of all the states in our ensemble. Any hope of doing so is futile: there would be no way of assessing this "representativeness." Rather, these are taken as 45 spatially limited views of the single parent microstate, treated as being infinitely far removed from each other in the crystal.

Following these simulations, instantaneous dislocation configurations were extracted at 0.075%, 0.15%, 0.225%, and 0.3% plastic strain. These extractions will allow us to examine how the correlation is affected as the simulation progresses and the total dislocation density rises. Fig. 2 shows the representative behavior of this collection of simulations. The stress-strain behavior is shown in (a) and the density behavior in (b). The black lines represent the mean behavior across all simulations, while the shaded region represents one standard deviation away from the mean. The plastic strain locations where configurations were extracted are shown on the x-axis.

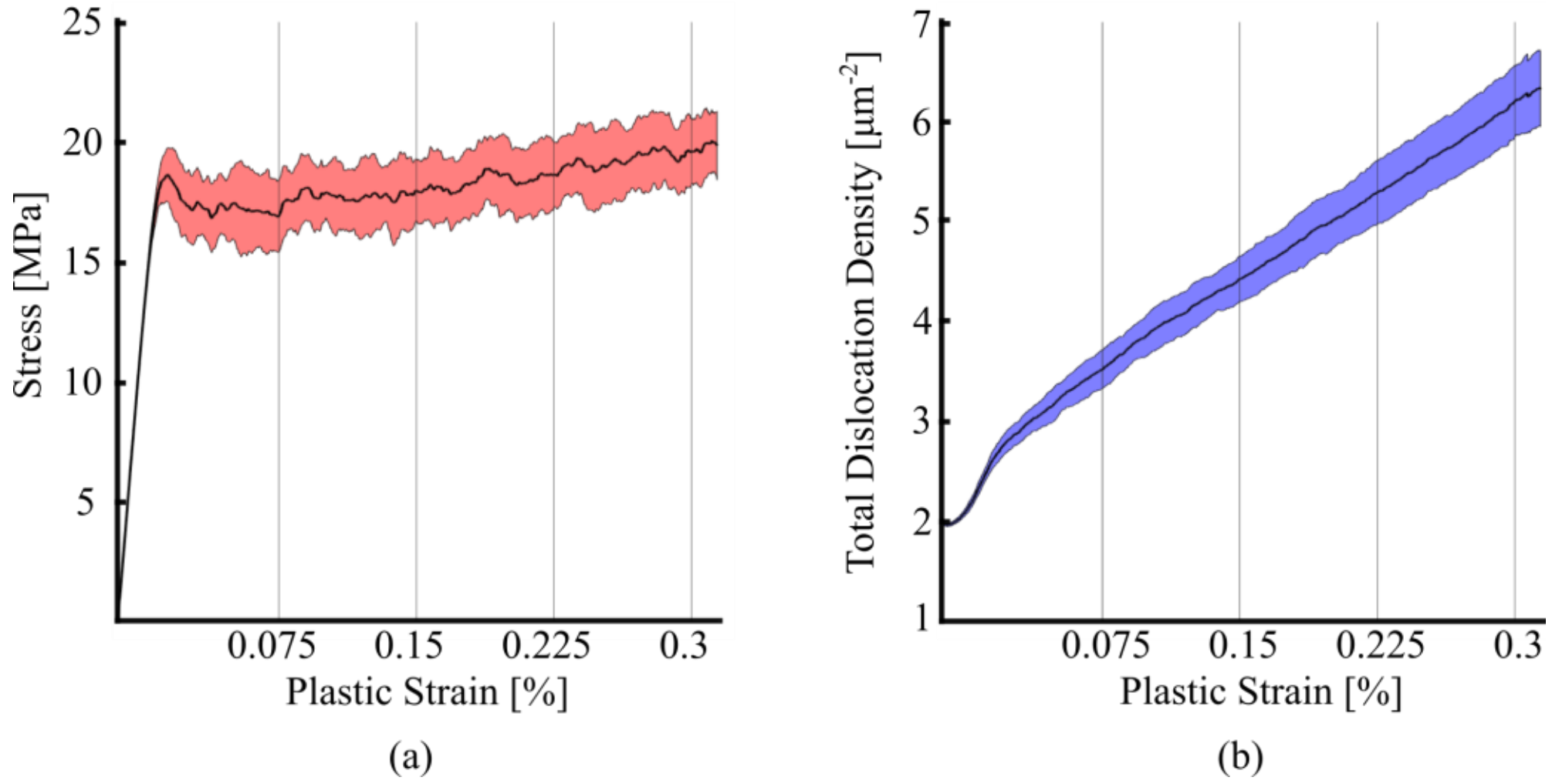

**Fig. 2** Representative simulation results from discrete dislocation dynamics. The representative stress-strain behavior of the 45 simulations is shown in (a), while the representative total dislocation density evolution is shown in (b). Both are plotted with respect to the total strain, while the plastic strain values where dislocation configurations were extracted are shown. Both show the mean behavior at each timestep of the 45 simulations in black, while the shaded region represents one standard deviation away from the mean. Figure (a) shows little strain hardening in the simulated regime. Figure (b) shows a linear increase in the total dislocation density, roughly doubling across the regions of plastic strain we consider.

**Table 1** Initial configuration parameters

| | | | | |
|---|---|---|---|---|
| Initial dislocation density | 2 $\mu m^{-2}$ | Box lengths: | [100] | 4.40 μm |
| Seed structure | Dipolar loops | | [010] | 4.87 μm |
| Length of segments in seed structure | 1 μm | | [001] | 5.84 μm |

**Table 2** Discrete dislocation dynamics parameters

| | | | |
|---|---|---|---|
| Strain rate | 20 $s^{-1}$ | Line tension model | de Wit |
| Time step | 2 ns | Relaxation | 200 ns (no reactions)<br>200 ns (reactions) |
| Temperature | 300 K | Lattice unit | 1.22 nm |
| Cross-slip | Activated | Slip plane distance (echelle) | 16.4 nm |

**Calculation of density fields**

The scheme for post-processing the dislocation configuration data to obtain correlation functions follows the line of reasoning resulting in equation (40). The crystal was discretized into an array of sample points 720 points long in the longest direction; this amounts to an 8.1 nm distance between points. The discrete-level convolution length $l_0$ was chosen to be twice this distance (16.2 nm) as this is the largest discretization distance in the simulation (the distance between discrete slip planes). Subsequent convolution length are multiples of the sample distance ranging from 24.3 nm to 162 nm.

All convolutions were performed using a cloud-in-cell weight function:

$$w_L(\boldsymbol{r}) = \begin{cases} \frac{1}{L^3}\prod_{i=1}^{3}\left(1-\frac{|r_i|}{L}\right) & \text{if } |r_i| < L \text{ for all } i = 1,2,3 \\ 0 & \text{otherwise} \end{cases}, \tag{46}$$

originally exposited in (Birdsall et al., 1969), and used previously in discrete-to-continuum treatments of dislocations in (Bertin, 2019) on account of its analytical solution for line integrals.

For each simulation output, we now have in hand the dislocation density $\rho$ and vector density $\boldsymbol{\rho}$ at all points in the crystal at 10 different levels of coarseness. Since the correlation calculation in equation (40) only involves the protocorrelation, and the support of the protocorrelation is the support of the "discrete" density, we only evaluate the higher convolutions on the support of $\tilde{\varrho}_i^*$.

Only glissile segments were used in the calculation of these densities: sessile junction segments (having Burgers vectors which are sums of the basic FCC Burgers vectors) were ignored.

## Computational Results

The main goal of the present work is to present a formulation which allows these dislocation correlation functions to be calculated. Preliminary results from this formulation will be shown results for a small (but important) class of correlations. While the free indices on $g^{(i,j)[\alpha,\beta]}(\boldsymbol{\Delta r})$ imply dependence on 3 vector components and 12 slip systems, we considered here the correlations for which $\alpha = \beta$. We will refer to such correlations as self-correlations. Since the two dislocation densities lie in the same slip system, we refer to all separation distances and density components in terms of a slip-system coordinate system consisting of the Burgers vector direction $\hat{b}$, the slip plane normal $\hat{n}$, and the binormal vector $\hat{a} \coloneqq \hat{n} \times \hat{b}$ (which we will refer to as the edge direction). Together these form a right-handed coordinate system $ban$. Separation vector components will be denoted as $\boldsymbol{\Delta r}_b, \boldsymbol{\Delta r_a}$, and $\boldsymbol{\Delta r}_n$, respectively. For all present intents and purposes, the density vector is a planar quantity, having only screw ($\hat{b}$) and edge ($\hat{a}$) components. The self-correlations between the screw-screw, screw-edge, and edge-edge components are discussed. The edge-screw component will not be discussed, as it is symmetric by parity to the screw-edge component.

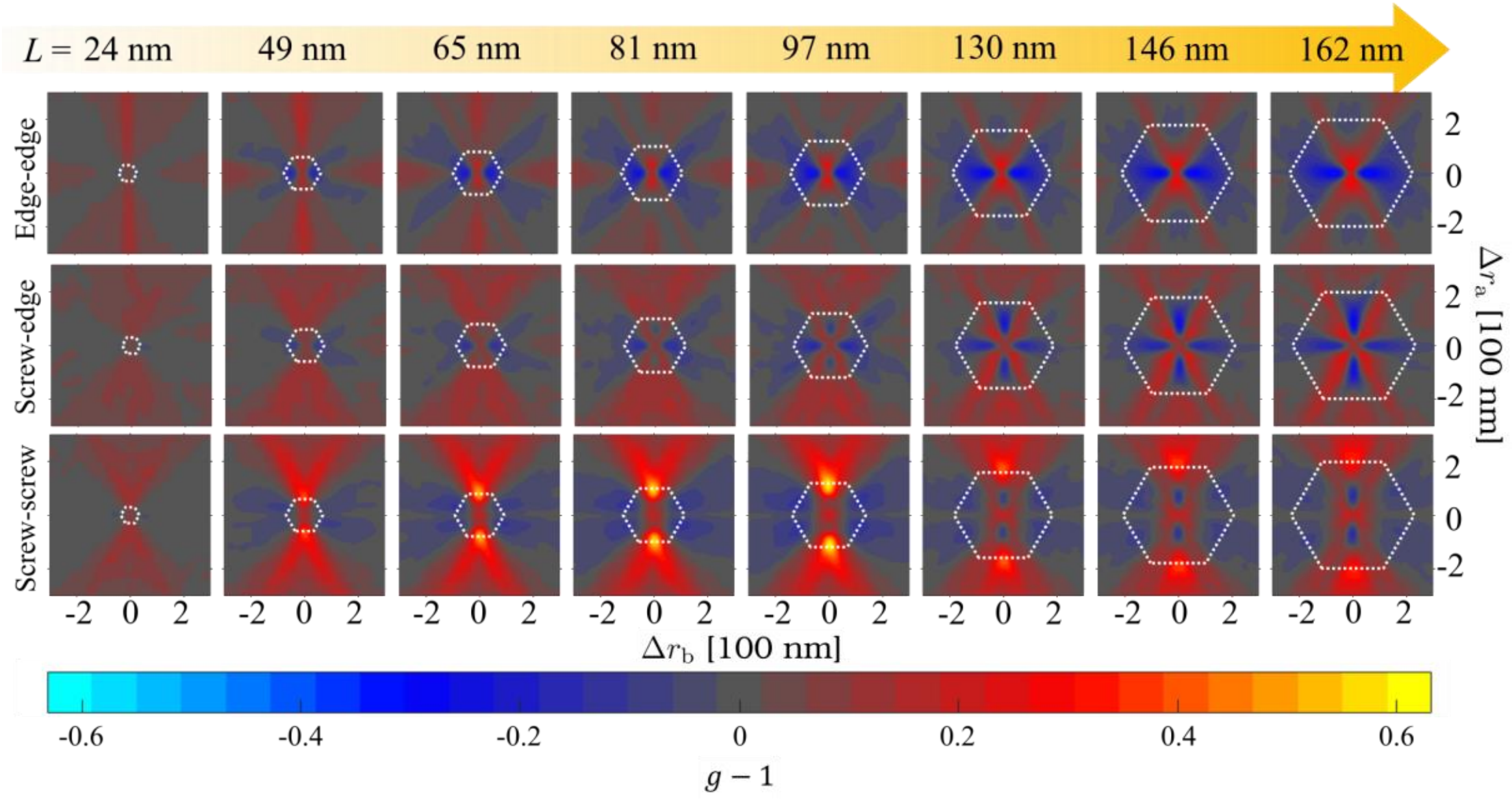

**Fig. 3** Dependence of the correlation tensor on the convolution length. All correlations shown are calculated from the 45 configurations at 0.30% strain, and the correlations are relative to a "discrete density" which is calculated with a 16 nm convolution length. The rows show the edge-edge, screw-edge, and screw-screw components of the correlation function and the columns demonstrate progressively longer correlation lengths. All figures show the correlation on the same color scale and relative to the same spatial dimensions. For reference, the border of the coarse averaging region (the support of $w_L$) is also shown as a hexagon, the intersection of the box with {111} type planes.

Among these results, we first present in Fig. 3 the dependence of the self-correlations on the convolution length $L$. These are all shown in the $\Delta r_n = 0$ plane (the slip plane itself). We quickly note that all the features of the correlation function seem to be relative to the convolution length. Past some minimal convolution length (≥65 nm), we see qualitatively similar spatial variation up to some spatial rescaling due to the convolution length. We suggest that the obscurity of some of the small features near the origin is not qualitatively different behavior, but rather arises due to the convolution on the order of 8.1 nm discussed in equation (26). For this reason, we choose the largest convolution length (162 nm) for the subsequent presentation of self-correlations, as it presents the clearest picture of the correlation, being less obscured by this effect.

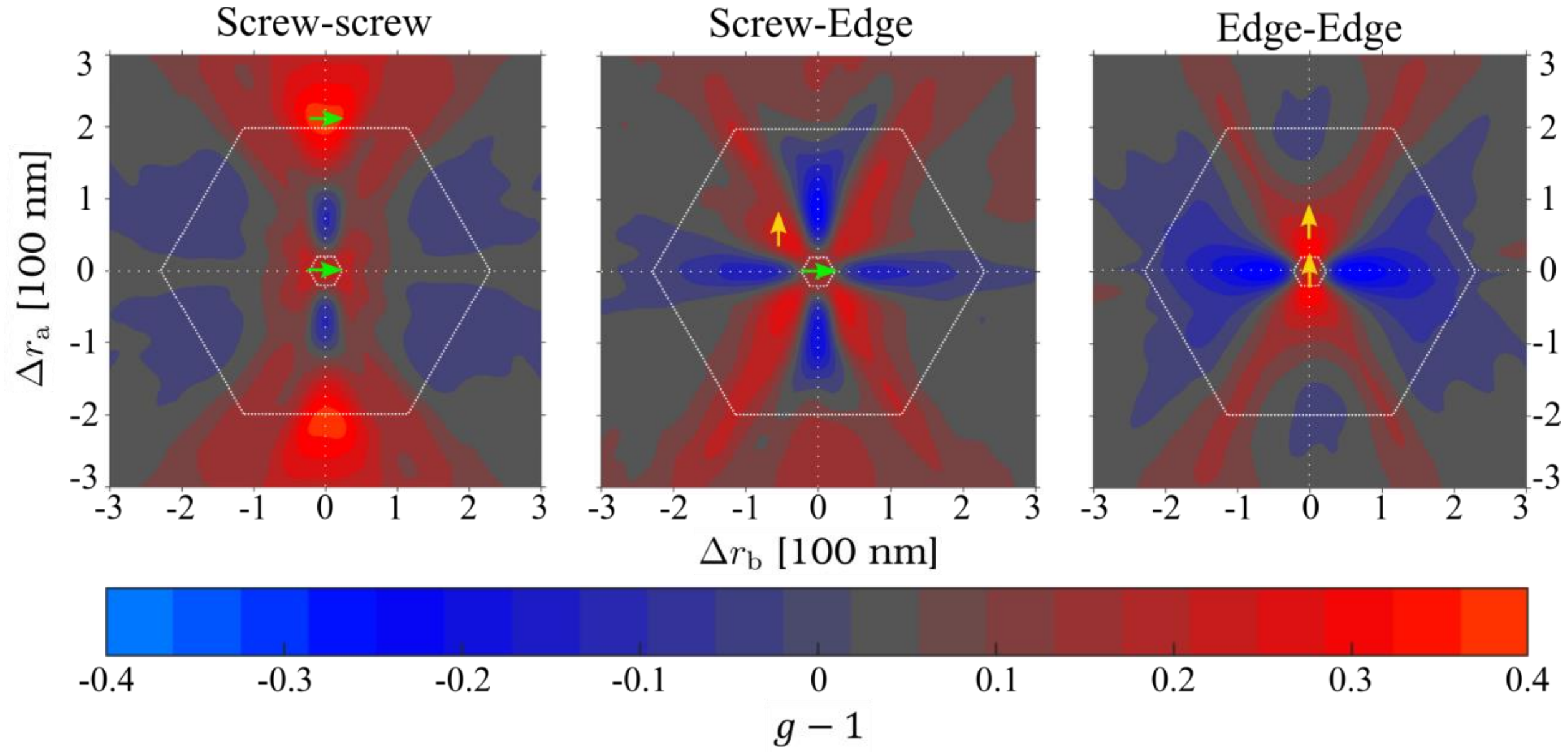

**Fig. 4** The spatial form of the correlation functions corresponding to the screw-screw, screw-edge, and edge-edge components. Shown are correlation functions calculated at 0.3% strain and a convolution length of 164 nm. The x-axis represents the spatial direction parallel to screw dislocations, and the y-axis represents the spatial direction parallel to edge dislocations. The red regions are those separation distances for which $\rho_i(\boldsymbol{r})\rho_j(\boldsymbol{r}+\boldsymbol{\Delta r})$ tends to underestimate the two-point density, and the blue regions are those where the two-point density is overestimated. This leads to a simple geometric interpretation in the case of the screw-screw and edge-edge components of the densities, but not so in the case of the screw-edge component. Rather than showing a simple circle with radius $L$ as in Fig. 3, we here show the intersection of the plane in view ({111} type) with the support of $w_L$ (a cartesian cube).

We would now like to discuss the spatial features of the correlation functions shown. Since all convolution lengths displayed similar spatial structures, we present the largest convolution length in Fig. 4 to allow resolution of the finer details. In order to interpret the spatial features, we return to what the correlation represents; namely, a component-wise recovery of the two-point density from the product of single-point densities:

$$\langle \varrho_i(\boldsymbol{r})\varrho_j(\boldsymbol{r}+\boldsymbol{\Delta r})\rangle = g^{(i,j)}(\boldsymbol{\Delta r})\rho_i(\boldsymbol{r})\rho_j(\boldsymbol{r}+\boldsymbol{\Delta r}). \tag{47}$$

An interpretation of spatial features can be garnered from this expression. If $g^{(i,j)}(\boldsymbol{\Delta r}) \approx 1$, $\rho_i(\boldsymbol{r})\rho_j(\boldsymbol{r}+\boldsymbol{\Delta r})$ (the naïve two-point density) represents an accurate picture of the two-point density at such separation distances. The naïve two-point density overpredicts the true two-point density at separation distances where $g^{(i,j)}(\boldsymbol{\Delta r}) < 1$ and underpredicts it at separations where $g^{(i,j)}(\boldsymbol{\Delta r}) > 1$. Since this is performed component-wise, it is naturally agnostic any geometric interpretation.

We note that it is dangerous to seek a geometric interpretation of the correlation functions in terms of dislocation lines. While much of the arrangement information encoded in the correlation functions deals with single-line effects (i.e. segments in some way connected to a differential segment at the origin), we caution against interpreting them in this geometric way due to complications in their meaning for dislocations of mixed character. However, if such an interpretation is desired, examine the pure screw and pure edge cases. Both the screw-screw and edge-edge correlations show a short-ranged underprediction in the "connected" direction ($\hat{b}$ and $\hat{a}$, respectively) and an overprediction in the "un-connected" direction ($\hat{a}$ and $\hat{b}$, respectively). However, in the case of pure screw densities, when the separation distance exceeds the convolution length, we see a marked underprediction of the screw density. In Fig. 4 this is on the order of 40%, in smaller convolution length systems in Fig. 3 this underprediction is on the order of 60%. It is worthy of note that these maxima are seen to be relative to the boundary of the coarse weight function $w_L$ (shown as a dotted line in the figures). This suggests that these maxima are not due to any geometric features of the dislocation configuration, but are rather caused by the averaging process. This will be discussed further when we examine the standard error of the calculations and the influence of the heterogeneity of the coarse $\rho_i(\boldsymbol{r})$.

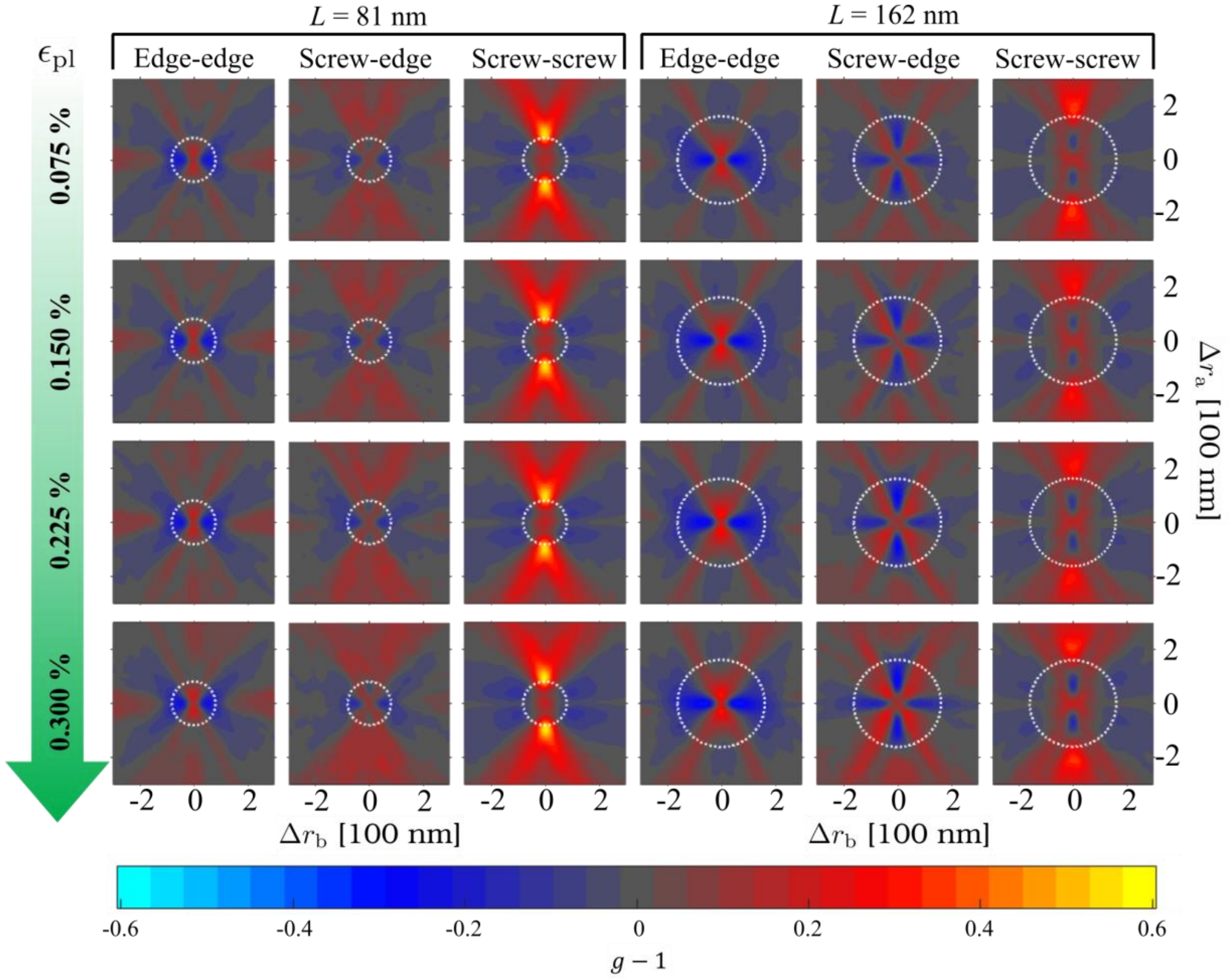

**Fig. 5** Dependence of the correlation tensor on the plastic strain (a surrogate for the dislocation density). This dependence (or rather lack thereof) is shown for two convolution lengths, 81 and 162 nm. The white dotted circle has radius equal to the convolution length $L$. Each component of the correlation tensor is shown for both convolution lengths. Plastic strain increases with descending row, with strain steps at every 0.075%

The second relation we would like to demonstrate is the dependence of the self-correlations on the plastic strain, shown in Fig. 5. Self-correlations were calculated separately from dislocation configurations at each strain step using 81 and 162 nm convolution lengths. In the course of the simulation (from 0.075% to 0.3%), the total dislocation density roughly doubles. However, we notice very little qualitative difference between these correlation functions as they run to higher strains.

Next we show the out-of-plane behavior of these self-correlation functions calculated at 0.3% strain with a convolution length of 162 nm. In Fig. 6, the first two columns show the coordinate planes passing through the origin (zero separation), i.e. $\Delta r_b = 0$ ($an$-plane) and $\Delta r_a = 0$ ($bn$- plane) planes, respectively. In these plots we notice marked anti-correlation ($g < 1$) for the off-plane separation vectors. This is expected, as we should expect the mean-field density to overpredict the two-point density at off-plane separation vectors on account of the likelihood of neighboring dislocations to be on the same slip plane as the dislocation at the origin. In the subsequent columns, slices parallel to the slip plane are shown for normal distances up to two times the convolution length, which equates to 324 nm in this case. We notice that this anti-correlation begins at distances as small as one tenth of the convolution length, and that the self-correlations converge to an uncorrelated state ($g = 1$) at distances greater than the convolution length.

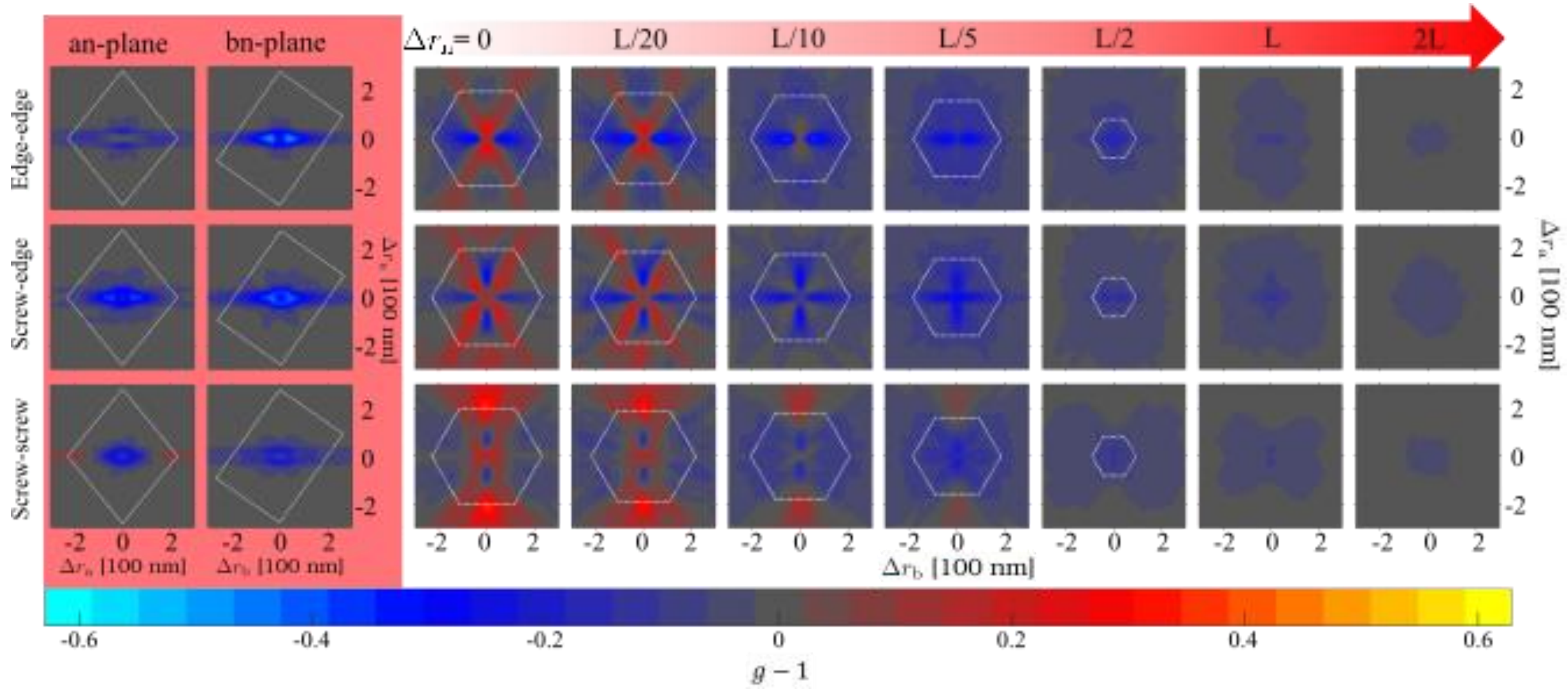

**Fig. 6** Out of plane behavior of the self-correlation calculated at 0.3% plastic strain and a convolution length of 162 nm. The first two columns show the behavior for separation distances in planes normal to the slip plane. Columns 3-7 show slices parallel to the slip plane at the indicated values of $\Delta r_{\rm n}$. Taken together, they show that the correlation function is largely relevant only on the slip plane itself, seeing slight anti-correlations for small out-of-plane separation vectors, and rapidly decaying to an uncorrelated state as $\Delta r_{\rm n} > L$ (the convolution length). Again, the intersection of the averaging volume with the viewing plane is shown in order to identify features which seem to be influenced by this boundary.

To examine this radial convergence to an uncorrelated state, we consider two types of radial correlation functions. The first is integrated over circles in the slip plane with radius $s$, while the second is integrated over the spherical surface with radius $r$:

$$g^{(i,j)}(s) \coloneqq \int d\theta \, g^{(i,j)}(s,\theta,n=0)\,, \tag{48}$$

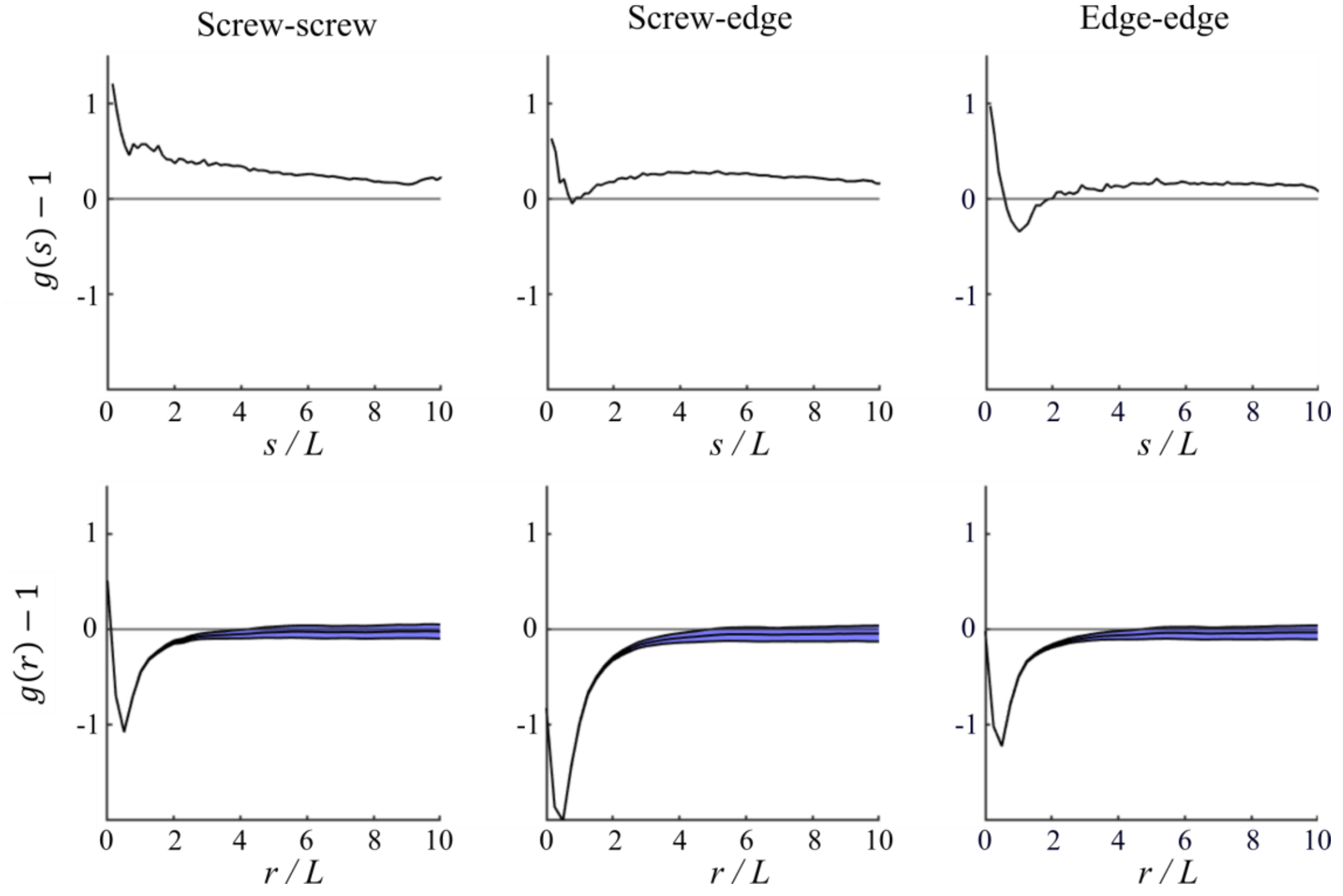

**Fig. 7** Radial correlation functions. Shown with ±1 standard error region. The correlations shown were calculated with a convolution length of 64 nm to examine large separation distances. The in-plane standard errors are too small to display on these axes. The in-plane radial correlation does not converge to an uncorrelated state ($g - 1 = 0$) in the spatial region examined. However, the spherical radial correlation function decays quickly to zero, with the error bars including zero by $r = 5L$.

$$g^{(i,j)}(r) := \int d\Omega \, g^{(i,j)}(r, \Omega) \,. \tag{49}$$

These are plotted in Fig. 7. The in-plane radial correlation function is not seen to converge to an uncorrelated state ($g - 1 = 0$). However, the spherical radial correlation function does converge to zero within 5 convolution lengths of the origin.

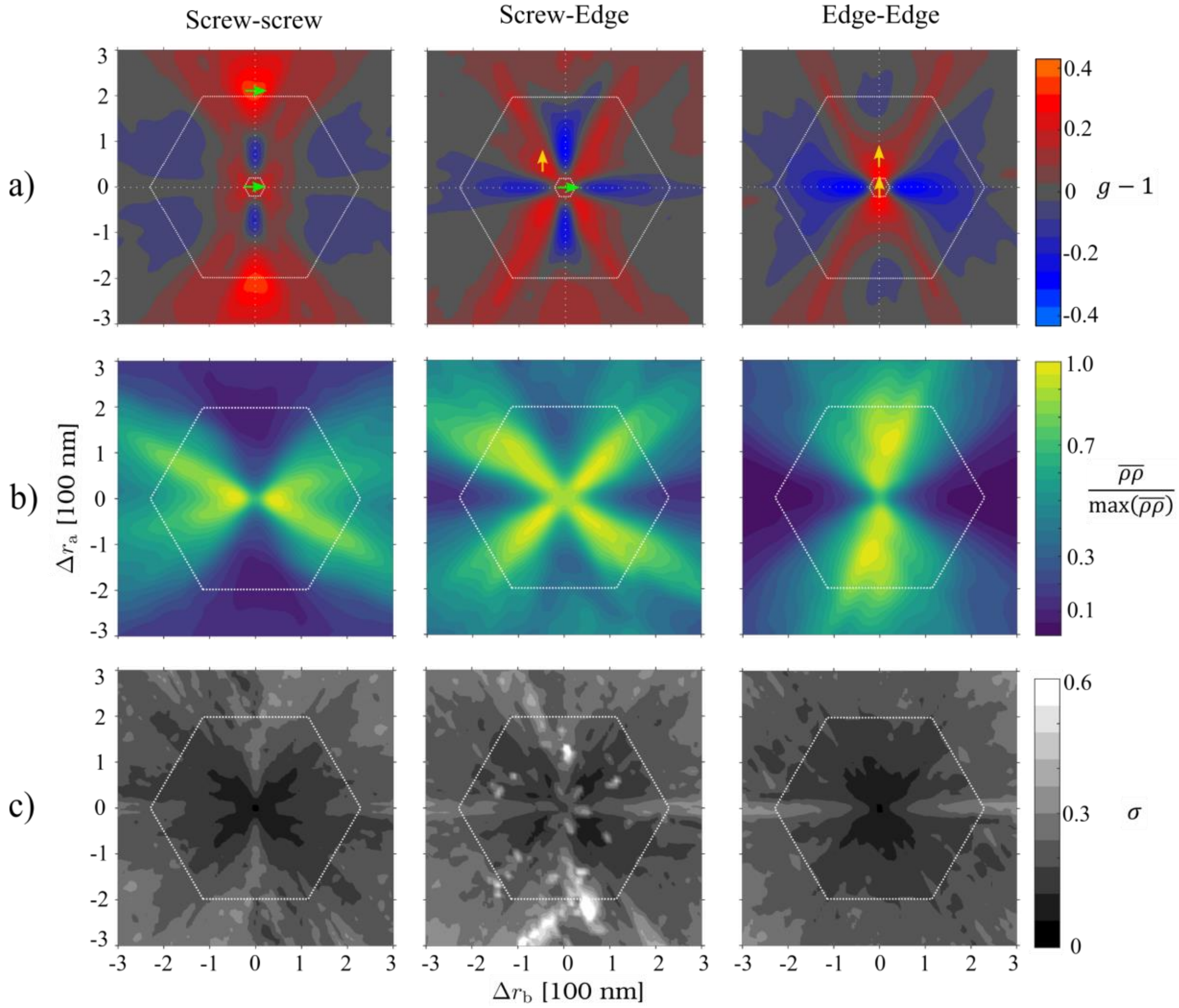

**Fig. 8** Quantities relevant to the certainty of the correlation calculations. For reference, the correlation functions are shown again in (a). The empirical mean of the coarse density product $\rho_i\rho_j$ normalized by the maximum value is shown in (b), and the standard error of the correlation calculation in (c). When the average density product (b) becomes small, this leads to a significantly increased standard error in the calculation, as it appears in the denominator of the correlation calculation of equations (40-41). Notably, several regions of the separation space on the plane are seen to have standard error values comparable to the magnitude of the correlation, namely in the unconnected directions (the y-axis in the screw-screw case and the x-axis in the edge-edge case). All quantities shown are calculated for the coarsest calculation ($L$ =162 nm).

Lastly, let us discuss the certainty of the calculation of the correlation functions. In Fig. 8b we see the effect of the heterogeneity of the coarse density field $\rho_i(\boldsymbol{r})$. At various separation distances, the average value of the coarse density product $\rho_i(\boldsymbol{r})\rho_j(\boldsymbol{r}+\boldsymbol{\Delta r})$ is seen to vary by roughly one order or magnitude. The effect of this is an increased standard error (Fig 8c) in these regions of low density product, calling into question the reliability of the calculation in these regions. Of particular note is the sharp increase in uncertainty in the unconnected directions in the screw-screw and edge-edge components of the correlation function. Beyond a distance of roughly 100 nm (5/8

L), the correlation value comes into question in the region very near the unconnected direction. Also of note is a marked increase in the standard error at the boundaries of the averaging volume. This belies a relationship to the averaging volume which depends not only the characteristic length $L$ but also on the weight function $w_L$ chosen.

## Discussion

In this section, we would like to discuss some of the implications of the tensorial form of the correlation used in the present work. Subsequently, we would like to discuss some of the preliminary implications of the findings with respect to incorporation into continuum dislocation dynamics schemes based on a vector density approach. This will be followed by a discussion of some open questions which were not settled by the present investigation.

### On the tensorial nature of the correlation

There is an important feature of the dislocation correlation function which warrants further discussion, namely, the tensor nature of the correlation. In the most basic sense, the correlation is simply a transformation between two tensorial quantities:

$$\langle \varrho_i(\boldsymbol{r})\varrho_j(\boldsymbol{r}')\rangle = g^{(i,j)}(\boldsymbol{r}-\boldsymbol{r}')\rho_i(\boldsymbol{r})\rho_j(\boldsymbol{r}'). \tag{50}$$

In the above equation and in all treatment throughout this work, the correlation $g^{(i,j)}$ represents a scalar transformation between one component of two tensor fields. However, one might have expected a more general linear form involving a fourth-rank tensor, namely:

$$\langle \varrho_i(\boldsymbol{r})\varrho_j(\boldsymbol{r}')\rangle = G_{ijkl}(\boldsymbol{r}-\boldsymbol{r}')\rho_k(\boldsymbol{r})\rho_l(\boldsymbol{r}')\,. \tag{51}$$

However, the form which we have used throughout is equivalent to a diagonal fourth-rank tensor of the following form:

$$G_{ijkl}(\boldsymbol{r}-\boldsymbol{r}') \coloneqq \sum_{m,n=1}^{3} g^{(m,n)}(\boldsymbol{r}-\boldsymbol{r}')\delta_{ikm}\delta_{jln}\,. \tag{52}$$

We have used the symbol $\delta_{ijk}$, which is equal to unity in the case where $i=j=k$ and zero otherwise. This rank three tensor—which is analogous to, but not to be confused with the Kronecker delta—allows diagonal second-rank tensors to be represented by a vector: $D_{ij} = \delta_{ijk}v_k$.

This diagonal form was chosen in the slip-system specific coordinate frame (*ban*, Burgers-edge-normal directions) not only to ensure the simplest component form (the density vector is planar in this case), but also because the interaction kernel is most simply expressed in a coordinate system containing the Burgers vector and the edge vector as well.

Under transformation of the underlying spaces by the same coordinate transformation $Q_{Ii}$, expression (51) transforms as follows:

$$\begin{aligned} \langle Q_{Ii} Q_{Jj} \varrho_i \varrho'_j \rangle &= Q_{Ii} Q_{Jj} \langle \varrho_i \varrho'_j \rangle \\ &= Q_{Ii} Q_{Jj} G_{ijkl} \rho_k \rho'_l \\ &= \left( Q_{Ii} Q_{Jj} G_{ijmn} Q^T_{mK} Q^T_{nL} \right) \left( Q_{Kk} Q_{Ll} \rho_k \rho'_l \right) \\ \langle \varrho_I \varrho'_J \rangle &= G_{IJKL} \rho_K \rho'_L. \end{aligned}$$

We thus see that $G_{ijkl}(\boldsymbol{r} - \boldsymbol{r}')$ transforms as a fourth rank two-point tensor (second rank in each leg). However, due to the two Kronecker deltas in the definition of $G_{ijkl}$, it has the same number of non-zero components as a second rank tensor. For simplicity, the non-zero components were referred to by their second rank equivalents throughout the work; $g^{(i,j)}$ was treated as having screw-screw, screw-edge, and edge-edge components, respectively (cf. Voigt notation, especially the diagonal components of his tensor bases). This choice of a diagonal form of the tensorial representation was decided in the local slip system coordinate frame (*ban*); if transformed to a different system the Kronecker delta form may not be preserved.

We further note that it was the diagonal form of the correlation tensor which allowed us to relate $\langle \varrho_i(\boldsymbol{r}) \varrho_j(\boldsymbol{r} + \boldsymbol{\Delta r}) \rangle$ and $\rho_i(\boldsymbol{r}) \rho_j(\boldsymbol{r} + \boldsymbol{\Delta r})$ by quasi-scalar operations. For example, examine the screw-edge component of the two-point density $\langle \varrho_s(\boldsymbol{r}) \varrho_e(\boldsymbol{r} + \boldsymbol{\Delta r}) \rangle$:

$$\begin{aligned} \langle \varrho_s(\boldsymbol{r}) \varrho_e(\boldsymbol{r} + \boldsymbol{\Delta r}) \rangle &= G_{sess} \rho_s(\boldsymbol{r}) \rho_s(\boldsymbol{r} + \boldsymbol{\Delta r}) + G_{sese} \rho_s(\boldsymbol{r}) \rho_e(\boldsymbol{r} + \boldsymbol{\Delta r}) \\ &\quad + G_{sees} \rho_e(\boldsymbol{r}) \rho_s(\boldsymbol{r} + \boldsymbol{\Delta r}) + G_{seee} \rho_e(\boldsymbol{r}) \rho_e(\boldsymbol{r} + \boldsymbol{\Delta r}) \end{aligned}$$

$$\langle \varrho_s(\boldsymbol{r}) \varrho_e(\boldsymbol{r} + \boldsymbol{\Delta r}) \rangle = g^{(s,e)} \rho_s(\boldsymbol{r}) \rho_e(\boldsymbol{r} + \boldsymbol{\Delta r}), \qquad (53)$$

allowing the diagonal components of the tensor to be expressed by scalar division of the corresponding components of the two density products. This division was then used to define a protocorrelation density which was central in the evaluation of the correlation.

### Incorporation in continuum dislocation dynamics

It was noted that the most significant non-spatial dependence of the self-correlation functions is on the convolution length $L$. This implies that any density-based continuum treatment of dislocations is dependent on the length-scale used to describe the system. The convolution length was used to force a distance of slow variation in the dislocation density field, and as a result has a close analog in the mesh size used to describe the dislocation density field in a spatially discretized continuum model. The findings imply that the correlation functions should be scaled with respect to that mesh size. The constant variation with respect to the mesh rather than the unscaled space will also allow for simpler integration in finite-element schemes, as they scale identically to the underlying shape functions.

Since it was observed that the self-correlation was stable with respect to the plastic strain (and as a result, total dislocation density), it is only necessary to supply a single form of the self-correlation fields at the beginning of a continuum simulation. It was initially a concern that these correlation functions would vary over the course of a simulation, greatly increasing the complexity as some sort of parallel simulation would have been needed to model the evolution of the correlation functions themselves. However, this does not seem to be the case. To incorporate these self-correlation functions, they must only be specified as a sort of initial condition to the simulation.

These two general considerations aside, we would like to speculate on how these self-correlations might be systematically incorporated into a continuum model. The most significant influence would involve a revision of the Peach-Koehler interactions of the dislocation densities. Treating this interaction as the conjugate configurational force to the dislocation density allows us to express this force as follows:

$$F_i(\boldsymbol{r}) = \frac{-\delta E}{\delta \rho_i(\boldsymbol{r})}, \tag{54}$$

$$F_i(\boldsymbol{r}) = -\int_{\mathcal{M}} \left(1 + \rho_i(\boldsymbol{r}) \frac{\delta g^{(i,j)}}{\delta \rho_i}\right) \rho_j(\boldsymbol{r}')\, \mathcal{E}_{ij}(\boldsymbol{r} - \boldsymbol{r}') g^{(i,j)}(\boldsymbol{r} - \boldsymbol{r}') d^3\boldsymbol{r}' \,. \tag{55}$$

If we assume that $\delta g^{(i,j)}/\delta \rho_i$ is vanishingly small, as we have already done in the statistical homogeneity assumption, we can neglect this second term. We assert that we can neglect this term, as part of the impetus of our formulation was to remove any relation between the local density field and the correlation. As a result we are left with a simple integration:

$$F_i(\boldsymbol{r}) = -\int_{\mathcal{M}} \rho_j(\boldsymbol{r}')\, \tilde{\mathcal{E}}_{ij}(\boldsymbol{r}-\boldsymbol{r}') d^3\boldsymbol{r}', \tag{56}$$

where we have incorporated the correlation as simply an alteration to the spatial dependence of the $ij$ interaction kernel:

$$\tilde{\mathcal{E}}_{ij}(\boldsymbol{\Delta r}) \coloneqq \mathcal{E}_{ij}(\boldsymbol{\Delta r}) g^{(i,j)}(\boldsymbol{\Delta r}). \tag{57}$$

The effects due to the correlated regions ($g - 1 \neq 0$) would introduce terms interpreted elsewhere as back and friction stresses. However, this energy kernel alteration circumvents the local density approximation (Zaiser, 2015) which underpins such back and friction stresses; such a local density assumption would be a poor approximation in our formulation given the significant variation of the correlation functions up to and past the convolution length. One may, however, perform the same expansion seen in (Zaiser, 2015) if higher-order terms are kept. This would result in an energy functional dependent not only on the local mean-field density and the integral of the correlation function, but also dependent on gradients of the density field and integral moments of the correlation.

**Future work**

Some questions regarding the self-correlation function remain open, and we would like to discuss them here. The present work did not establish upper bounds to any of the relations demonstrated. It was not shown whether the simple, linear scaling of the self-correlation with convolution length breaks down at large convolution lengths. It is also unknown whether the strain-independence of the self-correlation functions continues to hold at larger strains.

The dependences on convolution length and plastic strain may break down for the same reasons. The first possibility is that as the convolution length approaches the mean dislocation spacing ${\rho_0}^{-1/2}$ (~1.5 μm in the simulations presented, considering the average spacing between dislocations of like slip-system), the interactions being captured in the correlation functions are qualitatively different. Whereas at small convolution lengths the correlation functions capture line effects due to single dislocations, approaching the mean dislocation spacing will capture multi-dislocation effects such as dislocation patterning. We predict that upon convolution over lengths greater than this spacing, the correlations would become stable with respect to convolution length. As the strain increases, the length at which this transition might occur naturally decreases with the mean spacing. The second possible reason these relations could change at higher strains would be the introduction of lattice rotations. This would certainly affect the off-plane correlations as cross-slip becomes more common and slip planes are activated closer together. We can, however, assert that the relations demonstrated hold in the low-strain, low-convolution length regime which we have examined here. Discrete dislocation dynamics simulations might be run to slightly higher strain, but computation time needed to run a statistically significant number of such simulations would increase. Correlations at finite strains might necessitate other methods of investigation besides the discrete methods presented here.

Moreover, future work would be required to probe the reliability of this method of spatial empirical averages of the protocorrelation product. Sampling bias towards certain regions of the separation space were noted which are inherent to the line nature of the dislocation objects. More nuanced methods than the brute force calculation of the average two-point density presented here might be required, such as estimating the correlation function from stress or energy fluctuations within a discrete dislocation configuration.

A great deal is left to learn even in the present regimes of strain and convolution length by applying the formulation presented. For example, we have only considered the self-correlation of dislocation densities on the same slip system. The cross-correlations, on the other hand, contain information regarding the relative arrangement of different dislocation ‘species.’ This is especially important information which could inform corrections to dislocation reaction rates, which would in turn affect the strain hardening behavior of the crystal.

Additionally, the influence of the crystallography could introduce not only spatial symmetry breaking but also slip-system symmetry breaking. Efforts were made to suppress this dependence by averaging over the three strain directions as well as the twelve slip systems. However, a more nuanced investigation might reveal anisotropic effects due to the loading. On a similar note, it is unclear whether these correlation functions would change significantly for relaxed dislocation systems, as the configurations considered were all instantaneous snapshots of a dynamic simulation.

From a higher level, it would be interesting to examine alternative ensembles where different discrete fields are constrained. For example, one alternative to our line bundle ensemble would be to consider a high-order ensemble created by constraining the ensemble average either of lifted curves (Hochrainer, 2007) or of their moments in the angular space (Hochrainer, 2015). In lieu of the spatial convolution used herein, an operation analogous to that used in (Sandfeld et al., 2015) could be similarly used to evaluate correlation functions in this alternative ensemble.

## Conclusions

In this work, we have outlined a statistical framework which is implicit in any treatment of continuum dislocation dynamics employing density fields. It is our hope that this understanding will catalyze the definition of alternative ensembles, where different fields are constrained in their ensemble average. This allowed us to create a class of equivalent observations of random variables which one wishes to examine in average and to leverage basic statistical tools of empirical measurement to do so. It is our hope that this work will enable similar operations in other ensembles.

More specifically, we have in this work defined a line bundle ensemble by constraining a certain average of the discrete system to a spatial convolution of a singular dislocation density. Through the means just discussed, we were able to identify a set of observations which approach the correlation function in average.

This method was used to evaluate the three independent components of self-correlation function—by which we refer to the correlation between density components on the same slip system. These three independent components of the self-correlation functions were found to be strongly planar functions, with most of the interesting behavior being found at separation vectors falling in the slip plane. Moreover, the most significant factor affecting the form of the correlation function was found to be the convolution length used to define the mean field density: for lengths between 65 and 162 nm, the self-correlations are similar up to a rescaling of space proportional to the convolution length. No change in the correlation function was observed upon increase in total plastic strain—or equivalently, the total dislocation content of the simulated volume.

The implications which these findings have on continuum dislocation dynamics were discussed. It is the belief of the present authors that these correlation functions will provide an important correction to continuum dislocation dynamics models, introducing an altered form of the stress field and dislocation reaction rates. There are many features of these correlation functions which were beyond the scope of this particular document, but the results shown serve to demonstrate the validity of this approach to the calculation of correlation functions. It is our hope that this formulation will enable future studies of dislocation interactions in continuum dislocation models.

## Appendix 1: Physical Intuition of Set-theoretic Definition of Ensembles

To aid in understanding the process of defining an ensemble, we will treat the Gibbs' canonical ensemble in the set theoretic terms which have been presented. While this is an equilibrium system as opposed to our (highly) nonequilibrium system of interest, the intuition of spaces and level sets should be helpful nonetheless. Also, although these considerations are for systems of point particles, an understanding of the coarse-graining process will be helpful for understanding conceptually the dislocation ensemble.

### Spaces

Consider a system of $N$ particles, each with positions $\boldsymbol{r}_i \in \mathbb{R}^3$ and momenta $\boldsymbol{p}_i \in \mathbb{R}^3$. In this case, the microstate $\gamma$ of the system is a $6N$-tuple:

$$\gamma \coloneqq (\boldsymbol{r}_1, \dots, \boldsymbol{r}_N, \boldsymbol{p}_1, \dots, \boldsymbol{p}_N) \tag{A1}$$

and $\Gamma$ represents the space of all such microstates, commonly referred to as the phase space (Nolte, 2010):

$$\Gamma \coloneqq \{(\boldsymbol{r}_1, \dots, \boldsymbol{r}_N, \boldsymbol{p}_1, \dots, \boldsymbol{p}_N) : \boldsymbol{r}_i \in \mathbb{R}^3 \text{ and } \boldsymbol{p}_i \in \mathbb{R}^3 \text{ for all } 1 \le i \le N\} = \mathbb{R}^{6N}. \tag{A2}$$

Now we consider the quintessential coarse graining operation which involves level sets of the energy function. In this case, we consider as our coarse-graining function $\Psi = E(\boldsymbol{p}_1, \dots \boldsymbol{p}_N)$. Since this returns a scalar, our coarse-grained space $\mathrm{T}$ is simply $\mathbb{R}$. For clarity, let us examine the form of this map:

$$\Psi : \Gamma \to \mathrm{T} \qquad \Leftrightarrow \qquad E : \mathbb{R}^{6N} \to \mathbb{R} \tag{A3}$$

$$\Psi(\gamma \in \Gamma) \coloneqq \psi \in \mathrm{T} \qquad \Leftrightarrow \qquad E(\boldsymbol{r}_1, \dots, \boldsymbol{r}_N, \boldsymbol{p}_1, \dots, \boldsymbol{p}_N) \coloneqq \sum_{i=1}^{N} \frac{|\boldsymbol{p}_i|^2}{2m_i} \tag{A4}$$

The level sets of this map are referred to as macrostates $\Gamma_\epsilon$:

$$\Gamma_\epsilon \coloneqq \{\gamma \ : \ E(\gamma) = \epsilon\}. \tag{A5}$$

Notice that each of these macrostates represents a (hyper)sphere in the momentum portion of the space while extending as a cylinder in the position portion. These are all the equivalent configurations (microstates) which the particle system can occupy while retaining the same kinetic energy. There is a considerable amount of confusion about what such a macrostate map would look like in the case of dislocations, as the space is significantly more complex than $\mathbb{R}^{6N}$.

**Probabilities and Measures**

Now we wish to assign probabilities to subsets of Γ (and $\Gamma_\epsilon$). This requires two definitions: σ-algebras and measures. Broadly speaking, the former represents a “sufficiently large” collection of subsets of a given space, while the latter represents a map from these subsets to the real numbers.

A σ-algebra on a set $X$ is a collection of subsets of $X$ which: 1) contains $X$, 2) is closed under complement, and 3) is closed under countably infinite unions (Durrett, 2019). For an example of a σ-algebra, consider the Borel sets $\mathcal{B}$ on $\mathbb{R}$: the smallest σ-algebra containing the open sets. We may note that these are all sets which are easily imagined: they can be formed by unions of small intervals or of individual points. The definition of such a collection might seem obscure, but it is mostly useful to define measures.

A measure $\mu$ is a set function: it assigns non-negative real numbers to sets. The space of sets which is its domain is a σ-algebra, as it allows the statement of a measure’s defining (and most useful) property; for non-overlapping sets $A_i$ ($A_i \cap A_j = \emptyset$ for all $i \neq j$), the following decomposition must hold:

$$\mu\left(\bigcup_{i=1}^{\infty} A_i\right) = \sum_{i=1}^{\infty} \mu(A_i)\,. \tag{A6}$$

Intuitively, this implies that a set will retain its measure regardless of how many pieces we cut it into. This definition also carries two useful consequences: $A \subseteq B$ implies that $\mu(A) \leq \mu(B)$; $\mu(\emptyset) = 0$.

Two examples will be useful here. Firstly, consider the standard volume measure on the real numbers, referred to as the Lebesgue measure $\lambda$. This begins by simply assigning a map from all intervals to non-negative numbers:

$$\lambda\Big([a,b)\Big) = b - a \text{ for } b > a. \tag{A7}$$

This can be extended to the Borel sets by representing open sets as countable unions of such intervals and using the additive property from the definition of measures. The Lebesgue measure forms the theoretical basis of all real analysis, including but not limited to the rigorous definition of the integral.

The second important example is the probability measure. We will only here consider probability measures with density functions. Still considering measures on the real numbers, define a function $f\colon \mathbb{R} \to \mathbb{R}$ such that $\int_{\mathbb{R}} f(x)\, dx = 1$. We can then define the measure of a set $P(A)$ as:

$$P(A) = \int_A f(x)\, dx\,. \tag{A8}$$

This allows us to assign probabilities to sets. Notably, the fact that $P(\mathbb{R}) = 1$ allows us to measure complements as well:

$$\begin{aligned} P(A^C) + P(A) &= P(\mathbb{R}) \\ P(A^C) &= 1 - P(A) \end{aligned} \tag{A9}$$

Let us consider once again Gibbs' canonical ensemble. Representing again the states of constant energy as $\Gamma_\epsilon$, the fundamental assumption of statistical mechanics assigns all microstates of these level sets an equal probability. This allows us to define a probability measure as a map from a σ-algebra on $\Gamma_\epsilon$ to a non-negative number:

$$\begin{gathered} P_\epsilon\colon \sigma(\Gamma_\epsilon) \to [0,1] \\ P_\epsilon(A) \coloneqq \frac{\int_A d^{3N}\boldsymbol{r}\, d^{3N}\boldsymbol{p}}{\int_{\Gamma_\epsilon} d^{3N}\boldsymbol{r}\, d^{3N}\boldsymbol{p}} \end{gathered} \tag{A10}$$

If we use this probability measure, we are working in the microcanonical ensemble. Another tactic involves constraining the ensemble average of our macrostate variable. The canonical ensemble represents level sets of the temperature $\Gamma_T$ (where $T \coloneqq \langle E \rangle / k_B$), assigning probabilities:

$$P_T(A) = \frac{1}{Z(T)} \int_A e^{\frac{E(\boldsymbol{p})}{k_B T}} \, d^{3N}\boldsymbol{r} \, d^{3N}\boldsymbol{p}\,, \tag{A11}$$

where $Z(T) \coloneqq \int_{\Gamma_T} e^{\frac{E(\boldsymbol{p})}{k_B T}} \, d^{3N}\boldsymbol{r} \, d^{3N}\boldsymbol{p}$ denotes the canonical partition function and $k_B$ denotes the Boltzmann constant.

Defining this probability allows us to take averages as integration against this measure in the following way. The ensemble average of any quantity is now a function of the map $T$:

$$\langle A(\boldsymbol{p}, \boldsymbol{r}) \rangle (T) \coloneqq \frac{1}{Z(T)} \int_{\Gamma_T} A(\boldsymbol{r}, \boldsymbol{p}) \, e^{\frac{E(\boldsymbol{p})}{k_B T}} \, d^{3N}\boldsymbol{r} \, d^{3N}\boldsymbol{p}\,. \tag{A12}$$

Now, all of the necessary machinery for the set-theoretic definition of ensembles and coarse-graining has been shown for the canonical ensemble. For the present consideration of dislocations, we assume the existence of this machinery so as to define ensemble averages similar to that of equation (A12). Intuition from the form of this machinery is then used to identify a certain quantity (the protocorrelation density $\tilde{\varrho}(\boldsymbol{r})$) which can represent an independent measurement of the quantity being averaged.

## Appendix 2: A proof of use of the Radon-Nikodym theorem

In this appendix we present a proof of the validity of the operations resulting in equation (18). We will proceed in three stages. In the first, we will show that the measure formed by each component of the single point vector density is absolutely continuous with respect to the measure formed by the equivalent component of the mean-field density vector. We will then show that this implies absolute continuity of the 2-point discrete measure with respect to the product density measure. Lastly, we will show that the ensemble average preserves absolute continuity as well as the Radon-Nikodym derivative.

Before we begin, we will introduce some notation. A measure $\mu$ is absolutely continuous with respect to another measure $\nu$ if $\mu(A) = 0$ implies $\nu(A) = 0$. This is denoted $\nu \ll \mu$. A measure $\mu$ is singular with respect to another measure $\nu$ if there exists a set $A$ such that $\mu(A) = 0$ and $\nu(A^C) = 0$. This is denoted $\mu \perp \nu$. All sets defined will be of the Borel σ-algebra on $\mathcal{M}$ (the crystal space containing the dislocations) or the smallest σ-algebra on $\mathcal{M} \times \mathcal{M}$ containing the exterior product of all Borel sets in $\mathcal{M}$. All referenced theorems can be found in the appendix of (Durrett, 2019).

**Absolute continuity of the single point densities in a line bundle ensemble**

Let us examine the two measures formed component-wise by the single-point vector densities:

$$\mu(A \subseteq \mathcal{M}) = \int_A d^3\boldsymbol{r}\, \rho_i(\boldsymbol{r}) \tag{A13}$$

$$\nu_{\mathcal{L}}(A \subseteq \mathcal{M}) = \int_A d^3\boldsymbol{r}\, \varrho_i^{(\mathcal{L})}(\boldsymbol{r}) \tag{A14}$$

By Hahn's decomposition theorem (Durrett, 2019), let us use the signed measure $\mu$ to decompose $\mathcal{M}$ as $\mathcal{M}_+ \cup \mathcal{M}_-$, where $\mu_1(A_+) \geq 0$ for all $A_+ \subseteq \mathcal{M}_+$ and $\mu_1(A_-) \leq 0$ for all $A_- \subseteq M_-$, with $\mathcal{M}_+ \cap \mathcal{M}_- = \emptyset$. Similarly, decompose $\mathcal{M}$ into $\mathcal{M}_+^{\mathcal{L}}$ and $\mathcal{M}_-^{\mathcal{L}}$ by the measure $\nu_{\mathcal{L}}(A)$.

**Lemma A2.1:** The set of all dislocation microstates for which the sign of the $i$th component of their tangent vector is opposite to the sign of the $i$th component of the mean-field density vector, i.e.,

$$G_{\boldsymbol{r}} = \left\{\mathcal{L} \in \Gamma : \boldsymbol{r} \in (\mathcal{M}_+^{\mathcal{L}} \cap \mathcal{M}_-) \cup (\mathcal{M}_-^{\mathcal{L}} \cap \mathcal{M}_+)\right\} \tag{A15}$$

is of zero probability.

*Proof:* We will prove this by contradiction. Suppose $P(G_{\boldsymbol{r}}) > 0$. We note that the complement of $G_{\boldsymbol{r}}$ is

$$G_{\boldsymbol{r}}^C = \left\{\mathcal{L} \in \Gamma : \boldsymbol{r} \in (\mathcal{M}_+^{\mathcal{L}} \cap \mathcal{M}_+) \cup (\mathcal{M}_-^{\mathcal{L}} \cap \mathcal{M}_-)\right\}. \tag{A16}$$

Let us turn our eye to the following average:

$$\left\langle |\varrho_i(\boldsymbol{r})| \right\rangle - \left| \langle \varrho_i(\boldsymbol{r}) \rangle \right| \tag{A17}$$

This average is non-negative by Jensen's inequality. If we can show that it is strictly positive, we will have a contradiction of the line bundle requirement in equation (11). Suppose that $\boldsymbol{r} \in \mathcal{M}_+$ (the result for $\mathcal{M}_-$ is similar) and thus $\varrho_i$ is non-positive in $G_{\boldsymbol{r}}$ and non-negative in $G_{\boldsymbol{r}}^C$. This average then becomes

$$\left\langle |\varrho_i(\boldsymbol{r})| \right\rangle - \langle \varrho_i(\boldsymbol{r}) \rangle = \int_G dP(\mathcal{L}) |\varrho_i(\boldsymbol{r})| - \varrho_i(\boldsymbol{r}) + \int_{G^C} dP(\mathcal{L}) |\varrho_i(\boldsymbol{r})| - \varrho_i(\boldsymbol{r}) \tag{A18}$$

$$= -2\int_G dP(\mathcal{L})\, \varrho_i(\boldsymbol{r}) + \int_{G^C} dP(\mathcal{L})\, \varrho_i(\boldsymbol{r}) - \varrho_i(\boldsymbol{r}`)$$

$$= \ 2\int_G dP(\mathcal{L})\, |\varrho_i(\boldsymbol{r})|$$

Since the Hahn decomposition of $\mathcal{M}_{\pm}^{\mathcal{L}}$ is non-unique up to all sets in $\mathcal{M} \setminus \mathcal{L}$, we can choose $\mathcal{M}_{-}^{\mathcal{L}}$ such that $|\varrho_i(\boldsymbol{r})| > 0$ almost surely in $G_{\boldsymbol{r}}$. Thus, we obtain:

$$\langle|\varrho_i(\boldsymbol{r})|\rangle - \langle\varrho_i(\boldsymbol{r})\rangle = 2\int_G dP(\mathcal{L})|\varrho_i(\boldsymbol{r})|$$

$$> 2P(G_{\boldsymbol{r}}) \inf_{\mathcal{L}\in G_{\boldsymbol{r}}} |\varrho_i(\boldsymbol{r})|$$

$$> 0$$

This is a contradiction of the line bundle constraint. Therefore, the set of all dislocation microstates which point in the opposite direction to the mean-field density vector is of null probability, as desired.

We now note by Jordan's decomposition theorem (Durrett, 2019) that we now have $\mu(A) = \mu_+(A) - \mu_-(A)$ where $\mu_+(A) = \mu(A \cap \mathcal{M}_+)$ and $\mu_-(A) = \mu(A \cap \mathcal{M}_-)$, where $\mu_+ \perp \mu_-$. A similar operation can be performed for $\nu_{\mathcal{L}}(A) = \nu_{\mathcal{L}+}(A) - \nu_{\mathcal{L}-}(A)$. However, a corollary of the above lemma ensures additional singularity properties.

**Corollary:** $\mu_+$ $(\mu_-)$ is singular with respect to $\nu_-$ $(\nu_+)$ almost surely.

*Proof:* It follows from the lemma A2.1 that $\mu_+(\mathcal{M}_-) = 0$ and that $\nu_-(\mathcal{M}_+) = 0$ almost surely.

Lebesgue's decomposition theorem (Durrett, 2019) states that given any measures $\mu, \nu$ we can decompose $\mu$ as $\mu = \mu_{AC} + \mu_S$, where $\nu$ is absolutely continuous with respect to $\mu_{AC}$ and $\mu_S$ is singular with respect to $\nu$. This composition is unique up to a set of $\mu$-measure 0. The following lemma will leverage this decomposition to ensure absolute continuity of the single point positive and negative measures.

**Lemma A2.2:** $\mu_+$ $(\mu_-)$ is absolutely continuous with respect to $\nu_+$ $(\nu_-)$ almost surely.

*Proof:* By Lebesgue's decomposition theorem, we may define $\mu_1 = \mu_{\mathrm{AC}} + \mu_S$ uniquely up to a $\mu$-measure 0 set, where $\mu_{\mathrm{AC}}$ is absolutely continuous with respect to $\nu_+$ and $\mu_{\mathrm{S}}$ is singular with respect to $\nu_+$. However, we know that $\mu_-$ is singular with respect to $\nu_+$ almost surely by the above corollary. The uniqueness implies that $\mu_S = \mu_-$, and thus we arrive at $\mu_+$ absolutely continuous to $\nu_+$ almost surely. The same can be shown for $\mu_-$, $\nu_-$.

**Product measures**

We will now show that the product measures,

$$\mu_{\text{prod}}(A \subseteq \mathcal{M} \times \mathcal{M}) := \int_A d^3\boldsymbol{r}\, d^3\boldsymbol{r}' \rho_i(\boldsymbol{r})\rho_j(\boldsymbol{r}') \tag{A19}$$

$$\nu_{\mathcal{L}\times\mathcal{L}}(A \subseteq \mathcal{M} \times \mathcal{M}) := \int_A d^3\boldsymbol{r}\, d^3\boldsymbol{r}' \varrho_i(\boldsymbol{r})\varrho_j(\boldsymbol{r}')\,, \tag{A20}$$

inherit similar results as lemmas A2.1 and A2.2 from the single point measures.

**Lemma A2.2:** The set of all microstates for which the sign of the $ij$th component of their tangent vector product is opposite to the sign of the $ij$th component of the mean-field density vector product is of zero probability.

*Proof:* Decompose $\mathcal{M} \times \mathcal{M} := \Omega$ into $\Omega_+ = (\mathcal{M}_+ \times \mathcal{M}_+) \cup (\mathcal{M}_- \times \mathcal{M}_-)$ and $\Omega_- = (\mathcal{M}_- \times \mathcal{M}_+) \cup (\mathcal{M}_+ \times \mathcal{M}_-)$. Note that these unions are disjoint. The proof follows similarly to lemma A2.2.

**Lemma A2.4:** The Jordan decompositions $\mu_{\text{prod}} = \mu_{\text{prod}}^+ - \mu_{\text{prod}}^-$ and $\nu_{\mathcal{L}\times\mathcal{L}} = \nu_{\mathcal{L}\times\mathcal{L}}^+ - \nu_{\mathcal{L}\times\mathcal{L}}^-$ have the property $\nu_{\mathcal{L}\times\mathcal{L}}^+ \ll \mu_{\text{prod}}^+$ and $\nu_{\mathcal{L}\times\mathcal{L}}^- \ll \mu_{\text{prod}}^-$ almost surely.

*Proof:* The proto-measures on the collection of sets $A \times B$ where $A \in \mathcal{M}$ and $B \in \mathcal{M}$ defined as:

$$\mu_{\text{prod}}^+(A \times B) = \mu_+(A)\mu_+(B) + \mu_-(A)\mu_-(B)$$
$$\mu_{\text{prod}}^-(A \times B) = \mu_-(A)\mu_+(B) + \mu_+(A)\mu_-(B)$$

can be extended to measures on $\mathcal{M} \times \mathcal{M}$ satisfying the Jordan decomposition by using Carathéodory's extension theorem (Durrett, 2019). The mutual singularity property desired then is inherited from the single point densities (i.e. consider the sets $\Omega_+, \Omega_-$ from the proof of lemma A2.3).

**Effect of ensemble averaging**

**Theorem A2.1:** The measure $\nu_{\mathcal{L}\times\mathcal{L}}$ almost surely can be expressed in the following form:

$$\nu_{\mathcal{L}\times\mathcal{L}}(A) = \int_A \tilde{g}^{(i,j)}(\boldsymbol{r},\boldsymbol{r}')\, d\mu(d^3\boldsymbol{r}d^3\boldsymbol{r}') \tag{A21}$$

where

$$\tilde{g}^{(i,j)}(\boldsymbol{r},\boldsymbol{r}') = \frac{\varrho_i(\boldsymbol{r})\varrho_j(\boldsymbol{r}')}{\rho_i(\boldsymbol{r})\rho_j(\boldsymbol{r}')} = \tilde{\varrho}_i(\boldsymbol{r})\tilde{\varrho}_j(\boldsymbol{r}') \tag{A22}$$

*Proof:* The absolute continuity property $\nu^{+}_{\mathcal{L}\times\mathcal{L}} \ll \mu^{+}_{\text{prod}}$ and $\nu^{-}_{\mathcal{L}\times\mathcal{L}} \ll \mu^{-}_{\text{prod}}$ (lemma A2.4) implies the existence of the function $\tilde{g}^{(i,j)}(\boldsymbol{r},\boldsymbol{r}')$ by the Radon-Nikodym theorem. Since both measures are obtained by integration of a distribution in $\mathcal{M} \times \mathcal{M}$, the function is given by the scalar division of the two distributions.

We now have in hand the random variable $\tilde{g}^{(i,j)}$ which we have referred to in the text as the protocorrelation product. Only two things remain to be shown: a) that the average of the two-point density measure:

$$\nu_{\text{prod}}(A \subseteq \mathcal{M} \times \mathcal{M}) \coloneqq \int_A d^3\boldsymbol{r}d^3\boldsymbol{r}'\langle\varrho_i(\boldsymbol{r})\varrho_j(\boldsymbol{r}')\rangle \tag{A23}$$

is absolutely continuous with respect to the product measure $\mu_{\text{prod}}$ and b) that the Radon-Nikodym derivative of $\nu_{\text{prod}}$ with respect to $\mu_{\text{prod}}$ (the correlation function) is equal to the ensemble average of the protocorrelation product in theorem A2.1.

**Lemma A2.5:** If $\mu_{\text{prod}}(A) = 0$, $\nu_{\text{prod}}(A) = 0$, i.e. $\nu_{\text{prod}} \ll \mu_{\text{prod}}$.

*Proof:* It follows by the linearity of the integral that if a random variable $f = 0$ almost surely then $\langle f \rangle = 0$. It again requires decomposition into the positive and negative measures, but upon doing so one will still obtain that if $\mu_{\text{prod}}(A) = 0$, we have seen that $\nu_{\mathcal{L}\times\mathcal{L}}(A) = 0$ almost surely. If $\nu_{\mathcal{L}\times\mathcal{L}}(A) = 0$ almost surely, then $\nu_{\text{prod}}(A) = \langle\nu_{\mathcal{L}\times\mathcal{L}}(A)\rangle = 0$.

**Theorem A2.2:** $\nu_{\text{prod}}$ may be expressed as

$$\nu_{\text{prod}}(A) = \int_A g^{(i,j)}(\boldsymbol{r},\boldsymbol{r}')\, d\mu_{\text{prod}}(d^3\boldsymbol{r}d^3\boldsymbol{r}')$$

with

$$g^{(i,j)}(\boldsymbol{r},\boldsymbol{r}') = \langle \tilde{g}^{(i,j)}(\boldsymbol{r},\boldsymbol{r}') \rangle \tag{A24}$$

*Proof:* It follows from lemma A2.5 that $\nu_{\text{prod}}$ can be expressed with a Radon-Nikodym derivative as in the first equation, where

$$\begin{aligned}
\nu_{\text{prod}}(A) &= \int_A \frac{\langle \varrho_i(\boldsymbol{r})\varrho_j(\boldsymbol{r}')\rangle}{\rho_i(\boldsymbol{r})\rho_j(\boldsymbol{r}')}\, d\mu_{\text{prod}}(d^3\boldsymbol{r}d^3\boldsymbol{r}') \\
&= \int_A \frac{d\mu_{\text{prod}}(d^3\boldsymbol{r}d^3\boldsymbol{r}')}{\rho_i(\boldsymbol{r})\rho_j(\boldsymbol{r}')} \int_\Gamma dP\, \varrho_i(\boldsymbol{r})\varrho_j(\boldsymbol{r}') \\
&= \int_A d\mu_{\text{prod}}(d^3\boldsymbol{r}d^3\boldsymbol{r}') \int_\Gamma dP\, \frac{\varrho_i(\boldsymbol{r})\varrho_j(\boldsymbol{r}')}{\rho_i(\boldsymbol{r})\rho_j(\boldsymbol{r}')} \\
&= \int_A \langle \tilde{g}^{(i,j)}(\boldsymbol{r},\boldsymbol{r}')\rangle d\mu_{\text{prod}}(d^3\boldsymbol{r}d^3\boldsymbol{r}')
\end{aligned}$$

Since this is true for all sets $A \subseteq \mathcal{M} \times \mathcal{M}$, $g^{(i,j)}(\boldsymbol{r},\boldsymbol{r}') = \langle \tilde{g}^{(i,j)}(\boldsymbol{r},\boldsymbol{r}') \rangle$ on all sets that have $\mu_{\text{prod}} \neq 0$.

Thus we have demonstrated that any line bundle ensemble, not simply the convolution-based ensemble which we considered in the present work, can be described with correlations of the form defined in the subsection Product measures and correlation.

The influence of rotating the coordinate system is unclear. However, the safest way of proceeding, which does not interfere with the definition of the protocorrelation product is to assume $g^{(i,j)}$ represent the diagonal components of a fourth-rank tensor in a privileged coordinate frame where these are calculated. The most obvious choice of such a coordinate frame is the slip-system coordinate frame as it results in a planar dislocation density vector requiring fewer non-zero components of $g^{(i,j)}$ to describe the full correlation behavior.

## Abbreviations

3D: three-dimensional; 2D: two-dimensional.

**Availability of Data and Materials**

The collection of discrete dislocation dynamics simulations as well as the code used to generate the self-correlation functions can be accessed by contacting the corresponding author.

**Competing Interests**

The authors declare no competing interests.

**Funding**

This work is supported by the US Department of Energy, Office of Science, Division of Materials Sciences and Engineering, through award number DE-SC0017718 at Purdue University.

**Author's Contributions**

Both authors of this paper have contributed equally to the research presented in this paper. Both authors read and approved the final manuscript.

**Acknowledgments**

The authors used the microMegas code to perform the simulations used to collect the discrete dislocation configuration for correlation computation. The authors would like to acknowledge the referees, whose comments led us to make explicit assumptions we had implicitly made.

## References

K.-H. Anthony and A. Azirhi, Lagrangian field theory of plasticity and dislocation dynamics Attempts towards unification with thermodynamics of irreversible processes. Arch. Mech. **50**, 345 (1998).

M. Bao-Tong and C. Laird, Overview of fatigue behavior in copper single crystals-I. Surface morphology and stage I crack initiation sites for tests at constant strain amplitude. Acta Metall. **37**, 325 (1989).

N. Bertin, Connecting discrete and continuum dislocation mechanics: A non-singular spectral framework. Int. J. Plast. (2019).

C. K. Birdsall and D. Fuss, Clouds-in-clouds, clouds-in-cells physics for many-body plasma simulation. J. Comput. Phys. **3**, 494 (1969).

F. F. Csikor, I. Groma, T. Hochrainer, D. Weygand, and M. Zaiser, On the range of 3D dislocation pair correlations. Proc. 11th Int. Symp. Contin. Model. Discret. Syst. (2008).

J. Deng and A. El-Azab, Dislocation pair correlations from dislocation dynamics simulations. J. Comput. Mater. Des. **14**, 295 (2007).

B. Devincre, R. Madec, G. Monnet, S. Queyreau, R. Gatti, and L. Kubin, in Mech. Nano-Objects (2011), pp. 81–99.

R. Durrett, 5th ed. Probability: Theory and Examples, (Cambridge University Press, 2019).

A. El-Azab and G. Po, in Handb. Mater. Model. (2018), pp. 1–25.

I. Groma, Link between the microscopic and mesoscopic length-scale description of the collective behavior of dislocations. Phys. Rev. B **56**, 5807 (1997).

I. Groma and P. Balogh, Investigation of dislocation pattern formation in a two-dimensional self-consistent field approximation. Acta Mater. **47**, 3647 (1999).

I. Groma, F. F. Csikor, and M. Zaiser, Spatial correlations and higher-order gradient terms in a continuum description of dislocation dynamics. Acta Mater. **51**, 1271 (2003).

I. Groma, G. Györgyi, and B. Kocsis, Debye screening of dislocations. Phys. Rev. Lett. **96**, (2006).

A. N. Gulluoglu, D. J. Srolovitz, R. Lesar, and P. S. Lomdahl, Dislocation Distributions in Two Dimensions. Scr. Metall. **23**, 1347 (1988).

D. R. Hartree, The Wave Mechanics of an Atom with a Non-Coulomb Central Field Part I Theory and Methods. Math. Proc. Cambridge Philos. Soc. **24**, 89 (1928).

Hirth and Lothe, Theory of Dislocations, (John Wiley & Sons, 1982).

T. Hochrainer, Dissertation, Karlsruhe Institute of Technology, 2007.

T. Hochrainer, Multipole expansion of continuum dislocations dynamics in terms of alignment tensors. Philos. Mag. **95**, 1321 (2015).

T. Hochrainer, S. Sandfeld, M. Zaiser, and P. Gumbsch, Continuum dislocation dynamics: Towards a physical theory of crystal plasticity. J. Mech. Phys. Solids **63**, 167 (2014).

M. Kooiman, M. Hütter, and M. Geers, Effective mobility of dislocations from systematic coarse-graining. J. Stat. Mech. Theory Exp. **2015**, P06005 (2015).

E. Kröner, Benefits and shortcomings of the continuous theory of dislocations. Int. J. Solids Struct. **38**, 1115 (2001).

R. Lesar and J. M. Rickman, Incorporation of local structure in continuous dislocation theory. Phys. Rev. B **69**, 172105 (2004).

P. Li, S. X. Li, Z. G. Wang, and Z. F. Zhang, Unified factor controlling the dislocation evolution of fatigued face-centered cubic crystals. Acta Mater. **129**, 98 (2017).

P. Li, S. X. Li, Z. G. Wang, and Z. F. Zhang, Fundamental factors on formation mechanism of dislocation arrangements in cyclically deformed fcc single crystals. Prog. Mater. Sci. **56**, 328 (2011).

S. Limkumnerd and E. Van Der Giessen, Statistical approach to dislocation dynamics: From dislocation correlations to a multiple-slip continuum theory of plasticity. Phys. Rev. B **77**, 184111 (2008).

P. Lin and A. El-Azab, Implementation of annihilation and junction reactions in vector density-based continuum dislocation dynamics. Model. Simul. Mater. Sci. Eng. **28**, (2020).

H. C. Öttinger, Beyond equilibrium thermodynamics, (John Wiley & Sons, Hoboken, New Jersey, 2005).

H. C. Öttinger, in Beyond Equilib. Thermodyn. (John Wiley & Sons, Hoboken, New Jersey, 2005), pp. 213–260.

J. M. Rickman and R. Lesar, Issues in the coarse-graining of dislocation energetics and dynamics. Scr. Mater. **54**, 735 (2006).

S. Sandfeld, Dissertation, University of Edinburgh, 2010.

S. Sandfeld and G. Po, Microstructural comparison of the kinematics of discrete and continuum dislocations models. Model. Simul. Mater. Sci. Eng. **23**, (2015).

M. Sauzay and L. P. Kubin, Scaling laws for dislocation microstructures in monotonic and cyclic deformation of fcc metals. Prog. Mater. Sci. **56**, 725 (2011).

K. Starkey, G. Winther, and A. El-Azab, Theoretical development of continuum dislocation dynamics for finite-deformation crystal plasticity at the mesoscale. J. Mech. Phys. Solids **139**, 103926 (2020).

H. Stoyan and D. Stoyan, Simple stochastic models for the analysis of dislocation distributions. Phys. Status Solidi **97**, 163 (1986).

P.-L. Valdenaire, Y. Le Bouar, B. Appolaire, and A. Finel, Density-based crystal plasticity: From the discrete to the continuum. Phys. Rev. B **93**, 214111 (2016).

H. Y. Wang, R. Lesar, and J. M. Rickman, Analysis of dislocation microstructures: Impact of force truncation and slip systems. Philos. Mag. A **78**, 1195 (1997).

S. Xia, Dissertation, Purdue University, 2016.

S. Xia and A. El-Azab, Computational modelling of mesoscale dislocation patterning and plastic deformation of single crystals. Model. Simul. Mater. Sci. Eng. **23**, (2015).

S. Xia and A. El-Azab, A preliminary investigation of dislocation cell structure formation in metals using continuum dislocation dynamics. IOP Conf. Ser. Mater. Sci. Eng. **89**, (2015).

M. Zaiser, Local density approximation for the energy functional of three-dimensional dislocation systems. Phys. Rev. B **92**, 174120 (2015).

M. Zaiser, M. C. Miguel, and I. Groma, Statistical dynamics of dislocation systems: The influence of dislocation-dislocation correlations. Phys. Rev. B **64**, 2241021 (2001).